\documentclass[a4paper,11pt]{article}

\pdfoutput=1 

\usepackage{jheppub} 
                     
 \usepackage[T1]{fontenc} 

\usepackage{gensymb}
\usepackage{ae} 
\usepackage{aecompl}
\usepackage{bm} 
\usepackage[usenames,dvipsnames]{color}
\usepackage{amsmath}
\usepackage{amssymb}
\usepackage{epsfig}

\usepackage[utf8]{inputenc}

\usepackage{fontenc}
\usepackage{graphicx}
\usepackage{float}
\usepackage[lofdepth,lotdepth,caption=false]{subfig}
\usepackage{color}
\usepackage{booktabs}
\usepackage{tabularx}
\usepackage{multirow}
\usepackage{longtable}
\usepackage{appendix}
\usepackage{dcolumn}
\usepackage{rotating}%

\newcommand{\bi}{\begin{itemize}}
\newcommand{\ei}{\end{itemize}}

\def\beq{\begin{equation}}
\def\eeq{\end{equation}}

\newcommand{\bea}{\begin{eqnarray}}
\newcommand{\eea}{\end{eqnarray}}
\newcommand{\pmm}{P(\nu_\mu \rightarrow \nu_\mu)}
\newcommand{\pme}{P({\nu_{\mu} \rightarrow \nu_{e}})}
\newcommand{\pmt}{P({\nu_{\mu} \rightarrow \nu_{\tau}})}

\newcommand{\nue}{\nu_{\mu} \rightarrow \nu_{e}}

\newcommand{\numu}{{\nu_{\mu}} \rightarrow \nu_{\mu}}

\newcommand{\nutau}{{\nu_{\mu}} \rightarrow {\nu_{\tau}}}

\newcommand{\ta}{\theta_{12}}
\newcommand{\tb}{\theta_{13}}
\newcommand{\tc}{\theta_{23}}
\newcommand{\td}{\theta_{14}}
\newcommand{\te}{\theta_{24}}
\newcommand{\tf}{\theta_{34}}
\newcommand{\da}{\delta_{13}}
\newcommand{\db}{\delta_{24}}
\newcommand{\dc}{\delta_{34}}
\newcommand{\ldm}{\Delta m_{31}^2}
\newcommand{\sdm}{\Delta m_{21}^2}
\newcommand{\lldm}{\Delta m_{41}^2}

\newcommand{\chisq}{\Delta \chi^{2}}
\newcommand{\ie}{{\it i.e.}}
\newcommand{\eg}{{\it e.g.}}
\newcommand{\etc}{{\it etc.}}
\newcommand{\nova}{{NO$\nu$A}}

\title{
Exploring the new physics phases in 3+1 scenario in neutrino oscillation experiments
}

     \author[a,1]{Nishat Fiza}
     \author[b,c,2]{Mehedi Masud}
     \author[c,d,3]{Manimala Mitra}
     
     \affiliation[a]{Department of Physical Sciences, IISER Mohali, Knowledge City, SAS Nagar, Mohali - 140306, Punjab, India} 
     \affiliation[b]{Center for Theoretical Physics of the Universe, Institute for Basic Science (IBS), Daejeon 34126, Korea}
     \affiliation[c]{Institute of Physics, Sachivalaya Marg, Bhubaneswar, Pin-751005, Odisha, India}
     \affiliation[d]{Homi Bhabha National Institute, BARC Training School Complex, Anushakti Nagar, Mumbai-400094, India}
     
     \emailAdd{ph15039@iisermohali.ac.in}
     \emailAdd{masud@ibs.re.kr}
     \emailAdd{manimala@iopb.res.in}

\preprint{CTPU-PTC-21-02}
\abstract{The various global analyses of available neutrino oscillation data indicate the 
presence of the standard $3+0$ neutrino oscillation picture. 
However, there are a few short baseline anomalies that point to the possible existence 
of a fourth neutrino (with mass in the eV-scale), essentially sterile in nature. 
Should sterile neutrino exist in nature and its presence is not taken into consideration properly in the analyses of neutrino data, the interference terms arising due to the additional CP phases in presence of a sterile neutrino can severely impact the physics searches in long baseline (LBL) neutrino oscillation experiments. 
In the current work we consider one light (eV-scale) sterile neutrino and probe all the three CP phases ($\delta_{13}$, $\delta_{24}$, $\delta_{34}$) in the context of the upcoming 
Deep Underground Neutrino Experiment (DUNE) and also estimate how the results 
improve when data from NOvA, T2K and T2HK are added in the analysis.
We illustrate the $\chisq$ correlations of the CP phases among each other, and also with the three active-sterile mixing angles.  
Finally, we briefly illustrate how the relevant parameter spaces in the context of neutrinoless double beta decay get modified in light of the bounds in presence of a light sterile neutrino. }

\begin{document}
\maketitle
\flushbottom
\section{Introduction}
The successful discovery of the phenomena of neutrino oscillation~\cite{Fukuda:1998mi,Ahmad:2002jz} 
has led to an impressive amount of research with an aim to establish the 
standard 3-flavour (referred to as 3+0 hereafter) oscillation over a wide range of energy $(E)$ and neutrino propagation length $(L)$. 
In the commonly used Pontecorvo-Maki-Nakagawa-Sakata (PMNS) 
 parametrization~\cite{Zyla:2020zbs} of the leptonic mixing matrix, this requires fitting of the six standard oscillation parameters, 
 namely three mixing angles  ($\ta, \tb, \tc$), two mass squared differences ($\sdm = m_{2}^{2}-m_{1}^{2},  
 \ldm = m_{3}^{2}-m_{1}^{2}$) and one Dirac CP phase ($\da$). 
 A complete knowledge about the oscillation parameters will help to shed light on the following still 
 unresolved issues in the neutrino sector: whether there exists CP violation (CPV) in the 
 leptonic sector (\ie, whether $\da \neq 0, \pi$), whether the neutrino mass eigenstates 
 are arranged in normal ordering (NO, \ie, $\ldm > 0$) or inverted ordering (IO, \ie, $\ldm < 0$), and, 
 whether $\theta_{23}$ lies in the higher octant (HO, \ie, $\tc > \pi/4$) or in the lower octant 
 (LO, \ie, $\tc < \pi/4$). If leptonic CPV exists, it would provide a crucial missing ingredient~\cite{Sakharov:1967dj} in resolving another very fundamental elusive puzzle that is baryon asymmetry in the observed universe via a mechanism called 
 Leptogenesis~\cite{Fukugita:1986hr}. 
 Determination of mass ordering and $\tc$ octant will help in understanding 
 the origin of neutrino mass~\cite{Mohapatra:1979ia, Schechter:1980gr, Petcov:1982ya}, its Dirac/Majorana 
 nature via neutrinoless double beta decay~\cite{Haxton:1985am} and in exploring a new symmetry called $\mu-\tau$ symmetry~\cite{Lam:2001fb, Harrison:2002et}. 
 Presently running long baseline (LBL) experiments such as Tokai to Kamioka (T2K)~\cite{Abe:2013hdq} and NuMI Off-axis 
$\nu_e$ Appearance (NO${\nu}$A)~\cite{Ayres:2004js} have 
 started uncovering a few of  the open issues mentioned above.
Latest T2K results~\cite{Abe:2019vii} have been able to rule out a large 
range of values of $\da$ around $\pi/2$ at $3\sigma$ confidence level 
(C.L.) irrespective of mass ordering. It also excludes CP conservation ($\delta = 0$ or $\pi$) at 95\% C.L.   
NO$\nu$A, in its latest dataset~\cite{Acero:2019ksn} using both 
$\nu$ and $\bar{\nu}$ running mode hint towards NO at $1.9\sigma$ C.L. and shows a weak preference for 
$\theta_{23}$ lying in HO at a C.L. of $1.6\sigma$.  
Upcoming LBL experiments such as the Deep Underground Neutrino Experiment (DUNE)~\cite{Acciarri:2015uup, Abi:2020evt, Abi:2020qib}, Tokai to Hyper-Kamiokande (T2HK)~\cite{Abe:2015zbg}, Tokai to Hyper-Kamiokande with a second detector in Korea (T2HKK)~\cite{Abe:2016ero}, European Spallation Source $\nu$ Super Beam 
(ESS$\nu$SB)~\cite{Baussan:2013zcy} as well as future data from T2K and NO$\nu$A are expected to resolve the issues mentioned above with an unprecedented level of precision.\

Various global analyses~\cite{deSalas:2020pgw, globalfit, Capozzi:2018ubv, Esteban:2020cvm}, analyzing 
 neutrino data from a diverse array of sources such as atmosphere, particle accelerators, 
  sun and nuclear reactors have consolidated the effort to 
 build up the 3+0 oscillation picture.
 There are however, a few short-baseline (SBL) anomalies~\cite{Aguilar:2001ty, AguilarArevalo:2008rc, Mueller:2011nm, Mention:2011rk, Huber:2011wv} that hint towards the existence of oscillation 
 governed by $\mathcal{O}( \text{eV}^{2})$ mass squared difference ($\lldm = m_{4}^{2}-m_{1}^{2} \sim \text{eV}^{2}$) that cannot be accommodated 
 by the standard 3+0 scenario. 
 The effort to explain the SBL anomalies (excess of electron-like events at low energy) has led to the models with possible presence of a sterile, fourth type of neutrino (this scenario is referred to as 3+1 hereafter), which  
 can have small mixing with the three active neutrinos.
 This fourth generation of neutrino, if exists, has to be sterile since the LEP experiment~\cite{Decamp:1989tu} restricts the number of active neutrino flavours to three.
 In addition to the six standard oscillation parameters mentioned above, 3+1 scenario is 
 parametrized by the mass squared difference $\lldm$, three active-sterile mixing angles ($\td, \te, \tf$), 
 and two additional CP phases 
 which we refer as $\db$ and $\dc$.  
Recently the observation of a low energy electron-like event excess at a statistically significant $4.8\sigma$ by the Mini Booster 
Neutrino Experiment (MiniBooNE) in its latest dataset~\cite{Aguilar-Arevalo:2018gpe, Aguilar-Arevalo:2020nvw} has further boosted the search for sterile neutrinos. 
Various other existing and future facilities aim to search for sterile neutrinos with high precision using different neutrino sources and detection techniques.  
These facilities include IceCube~\cite{Aartsen:2020fwb}, Karlsruhe Tritium Neutrino Experiment (KATRIN)~\cite{Osipowicz:2001sq}, FermiLab's Short Baseline Neutrino (SBN) programme~\cite{Antonello:2015lea}, ANTARES~\cite{Collaboration:2011nsa}, Neutrino Experiment for Oscillation at Short baseline (NEOS)~\cite{Ko:2016owz}, Short baseline neutrino Oscillations with a novel Lithium-6 composite scintillator Detector(SoLid)~\cite{Abreu:2017bpe}, 
Neutrino-4~\cite{Serebrov:2018vdw}, Precision Reactor 
Oscillation and SPECTrum Experiment (PROSPECT)~\cite{Ashenfelter:2018iov}, Sterile Reactor Neutrino Oscillations (STEREO)~\cite{Allemandou:2018vwb, AlmazanMolina:2019qul}, 
Detector of the reactor AntiNeutrino based on Solid Scintillator (DANSS)~\cite{Alekseev:2018efk}, 
J-PARC Sterile Neutrino Search at J-PARC Spallation Neutron Source($\text{JSNS}^{2}$)~\cite{Rott:2020duk, jsns_nu2020}.  
Neutrino-4 has recently claimed to observe 
active-sterile oscillation at $3.5\sigma$ around the vicinity of $\lldm \approx 7 \text{ eV}^{2}$ and $\td \approx 18^{\circ}$ by analyzing the its reactor antineutrino data accumulated since 2016~\cite{Serebrov:2020rhy, Serebrov:2020kmd, neutrino4_nu2020}. 
Though it has later been argued in literature that considering a more appropriate log-likelihood distribution~\cite{Giunti:2020uhv, Coloma:2020ajw} or correct energy resolution~\cite{Giunti:2021iti}, the statistical significance of the active-sterile observation results in reactor neutrino experiments actually come down to a much lower value. 
ANTARES with its 10 years of 
data have found very mild ($\sim 1.6 \sigma$) signature of 
sterile neutrino in their analyses~\cite{Albert:2018mnz}. 
Experiments looking for Coherent Elastic Neutrino-Nucleus Scattering (CE$\nu$NS)~\cite{Akimov:2017ade} can also act as very 
useful probes of sterile neutrinos, as shown by the authors of \cite{Miranda:2020syh}.

The effect of sterile neutrino, should it exist with $\lldm \sim 1\text{ eV}^{2}$, is most pronounced around $L/E \sim 1$ 
km/GeV\cite{Gandhi:2015xza, Boser:2019rta}. 
In LBL experiments where $L/E \sim 500$ km/GeV at the far detectors (FD), the high frequency induced by 
$\lldm \sim 1\text{ eV}^{2}$ gets averaged out due to the finite energy resolution of the detector. 
Nevertheless, it has been shown~\cite{Klop:2014ima, Berryman:2015nua, Gandhi:2015xza, Palazzo:2015gja, Agarwalla:2016mrc, Agarwalla:2016xxa, Agarwalla:2016xlg, Dutta:2016glq, Rout:2017udo, Kelly:2017kch, Ghosh:2017atj, 
Choubey:2017cba, Coloma:2017ptb, Tang:2017khg, Choubey:2017ppj, Agarwalla:2018nlx, Gupta:2018qsv, Choubey:2018kqq, 
deGouvea:2019ozk, Ghoshal:2019pab, Agarwalla:2019zex, Majhi:2019hdj, Ghosh:2019zvl, Chatterjee:2020yak} that even at the FDs of these LBL experiments, the interference effects provided 
by the additional CP phases play very significant roles in spoiling the sensitivities to the crucial issues of CPV, MH 
and $\tc$ octant\footnote{For a recent comprehensive status report of the impact of a light sterile 
neutrino in probing these issues at LBL see, for \eg, \cite{Giunti:2019aiy, Diaz:2019fwt, Palazzo:2020tye} and the references therein.}. 
For instance, as shown in \cite{Dutta:2016glq}, the CPV sensitivity becomes a wide band whose width depends on the unknown magnitudes of the 
sterile phases $\db$ and $\dc$, - leading to serious confusion in interpreting the results as CP violation 
or CP conservation.  
The constraints on the active-sterile mixing angles $\theta_{i4}$ ($i=1, 2, 3$) do of course reduce such ambiguities  
in the interpretation of the results to some extent. 
But, a clear idea about how the LBL experiments are able to measure the sterile phases $\db$ and $\dc$ given 
the constraints on the active-sterile mixing angles, will certainly minimise the obfuscation when their data 
are analyzed with a view to tackle the unresolved physics issues. 
The issue of measurement of one sterile phase ($\db$ or $\delta_{14}$, depending on the 
parametrization used) has been addressed to some extent in 
literature. 
 \cite{Dutta:2016glq} analyzes how the sterile phases impact the standard CPV 
and mass ordering measurements and also probe the joint parameter space $\da-\db$ for DUNE.  
The authors of \cite{Agarwalla:2016xxa} use a slightly different parameterization and discuss the CPV arising from the individual sterile CP phases and probe the joint parameter space $\da-\delta_{14}$ at DUNE. 
They further extend their analysis by combining simulated data from other LBL experiments 
such as T2HK~\cite{Agarwalla:2018nlx} and ESS$\nu$SB~\cite{Agarwalla:2019zex}. 
Exploring the parameter space of $\da-\db$ has also been addressed in \cite{Choubey:2017ppj} for DUNE, T2HK, T2HKK and their combinations. 
The authors of \cite{Choubey:2017ppj} further illustrate how the individual phase $\db$ 
can be measured by these experiments at various C.L. by assuming four possible true 
values ($0, \pm \pi/2, \pi$). 
\cite{Chatterjee:2020yak} shows how the recent data from T2K and 
NO$\nu$A can help to probe the parameter space of $\da-\delta_{14}$. 
Most recently, the authors of \cite{Chatla:2020bqb} estimate how the difference ($\delta_{14}-\delta_{24}$) in the sterile phases can impact in constraining the standard 
$\delta_{13}-\theta_{23}$ parameter space.

It is noteworthy that the sterile CP phase $\dc$ and its correlation with the other phases 
has been little addressed in literature.  
In the present manuscript we tackle the very relevant issue of estimating the capability to reconstruct 
all three CP phases ($\da, \db, \dc$), taking into consideration their $\chisq$ correlations with each other 
and also with the active-sterile mixing angles ($\td, \te, \tf$). 
We carry out this exercise in the context of DUNE and illustrate the improvement when combined with T2K, NO$\nu$A (both these currently running experiments are simulated upto their present exposure) and T2HK. 
Apart from studying these CP phases in detail, another crucial aspect in which our analysis differs from the existing studies mentioned above is that we have taken into consideration the current $3\sigma$ hint of CP violation and the corresponding exclusion region of the CP phase $\da$ by T2K data~\cite{Abe:2019vii}. 
Moreover, in addition to illustrating how the 2-d parameter spaces for the CP phases can 
be probed, we also analyze how the individual CP phases can be reconstructed (after marginalizing all other relevant parameters) by the 
experiments, irrespective of the actual value they might have in nature. 
Additionally we have also considered the $\nutau$ channel (in addition to $\nue$ and $\numu$) in our 
study and estimated the capability of the projected data to measure all three CP phases in detail. 
This enables us to probe the parameter spaces associated to $\tf$ and $\dc$ with 
better sensitivity. 
 
The present manuscript is organised as follows. 
In Sec.\ \ref{sec:basics} we give a brief account of the constraints on the 
sterile neutrino parameters and how the CP phases affect the relevant probabilities. 
In  Sec.\ \ref{sec:analysis} we describe the methodology of our statistical analyses. 
In Sec.\ \ref{sec:phase_phase} we discuss the $\chisq$ correlations among various CP phases, taking a pair of phases at a time. 
We also discuss the potential to reconstruct the three CP phases for all possible true 
values by performing simulations of DUNE, T2K, NOvA and T2HK. 
The role of individual oscillation channels in such reconstructions are analyzed in 
Sec.\ \ref{sec:channel}.  
We estimate the $\chisq$ correlations among all the CP phases and the active-sterile 
mixing angles in Sec.\ \ref{sec:phase_angle}. 
Finally in Sec.\ \ref{sec:ndbd} we briefly discuss how the relevant parameter spaces 
associated to Neutrinoless Double Beta Decay gets modified in light of the constraints 
on one eV scale sterile neutrino, followed by conclusion.  

 \section{Basics}                                                                     
 \label{sec:basics}
We first discuss the oscillation probabilities for the three channels ($\pme$, $\pmm$ and $\pmt$) in 3+1 scenario. 
Since the expressions become immensely complicated in matter, we show them in vacuum 
and these will act as useful templates for gaining the physics insights in explaining our 
subsequent sensitivity results. 
For the mixing matrix we follow the parametrization scheme adopted in \cite{Gandhi:2015xza}:
\begin{equation}
U^{3+1} = R(\tf, \dc) R(\te, \db) R(\td) R(\tb) R(\tc, \da) R(\ta), 
\label{eq:param}
\end{equation} 
where $R(\theta_{ij}, \delta_{ij})$ is a rotation in the $ij-$th plane with an associated phase 
$\delta_{ij}$ such that, for \eg,
\begin{equation}
R(\tf, \dc) = \left( \begin{array}{cccc}
1 & 0 & 0 & 0 \\ 0 & 1 & 0 & 0 \\ 
0 & 0 & \cos\tf & e^{-i\dc}\sin\tf \\ 0 & 0 & -e^{i\dc}\sin\tf & \cos\tf
\end{array}
\right). 
\label{eq:rotation}
\end{equation}
Now we discuss briefly on the allowed ranges of the sterile sector parameters, as estimated in great detail in the global analysis of various neutrino data~\cite{Dentler:2018sju}. 
$|U_{e4}|^{2}$ is bounded by $\nu_{e}$ and $\bar{\nu}_{e}$ disappearance searches and is equal 
to $\sin^{2}\td$. 
The combined atmospheric neutrino data from IceCube, DeepCore and SK, at $99\%$ C.L. (2 DOF) put 
the bound $|U_{e4}|^{2} \lesssim 0.1$. This implies $\theta_{14} \lesssim 18.4^{\circ}$. 
It can be seen from all $\nu_{e}$ and $\bar{\nu}_{e}$ searches that the best fit value of $|U_{e4}|^{2}$ 
is approximately equal to $0.01$, which gives $\td \approx 5.7^{\circ}$. 
The data from $\nu_{\mu}$ and $\bar{\nu}_{\mu}$ disappearance searches put the following $99\%$ C.L. (2 D.O.F.) constraints: $|U_{\mu4}|^{2} \lesssim 0.01$ and $|U_{\tau4}|^{2} \lesssim 0.17$. 
Since, in our parametrization $|U_{\mu4}| = \cos \td \sin \te$ and $|U_{\tau 4}| = \cos \td \cos \te \sin \tf$, 
the corresponding bounds on $\te$ and $\tf$ can easily be translated to 
$\te \lesssim 6.05^{\circ}$ and $\tf \lesssim 25.8^{\circ}$. 
The allowed values of $\lldm$ roughly lie in the range of $1-10 \text{ eV}^{2}$ 
and we consider it to be $1.3 \text{ eV}^{2}$ in the present work as per the global analysis\footnote{In our notation, $m_{1,2,3}$ are the masses of the three active neutrinos, and $m_4$ denotes the mass of the sterile neutrino. Also, $\Delta m^2_{ij} = m^2_i - m^2_j$.}.

Using the standard approach for deriving oscillation probability, we obtain for the $\nu_{\alpha} 
\to \nu_{\beta}$ ($\alpha, \beta = e, \mu, \tau, s$ and $\alpha \ne \beta$) transition probability, 
\begin{align}
\label{eq:pab}
P_{\alpha\beta}^{3+1} &=  4|U_{\alpha 4}U_{\beta 4}|^{2} \times 0.5 \nonumber \\
&-4\text{Re}(U_{\alpha 1} U_{\beta 1}^{*} U_{\alpha 2}^{*} U_{\beta 2})\sin^{2}\Delta_{21}
+2\text{Im}(U_{\alpha 1} U_{\beta 1}^{*} U_{\alpha 2}^{*} U_{\beta 2})\sin2\Delta_{21} \nonumber \\
&-4\text{Re}(U_{\alpha 1} U_{\beta 1}^{*} U_{\alpha 3}^{*} U_{\beta 3})\sin^{2}\Delta_{31}
+2\text{Im}(U_{\alpha 1} U_{\beta 1}^{*} U_{\alpha 3}^{*} U_{\beta 3})\sin2\Delta_{31} \nonumber \\
&-4\text{Re}(U_{\alpha 2} U_{\beta 2}^{*} U_{\alpha 3}^{*} U_{\beta 3})\sin^{2}\Delta_{32}
+2\text{Im}(U_{\alpha 2} U_{\beta 2}^{*} U_{\alpha 3}^{*} U_{\beta 3})\sin2\Delta_{32},
\end{align}
where $\Delta_{ij} = \frac{\Delta m^{2}_{ij} L}{4E}.$
In deriving Eq.\ \ref{eq:pab}, we have used the unitarity of the $4 \times 4$ mixing matrix $U$ and 
applied the usual assumptions that the term containing mass square splitting between $m_4$ and $m_i\ (i = 1, 2, 3)$, \ie,  $\sin^{2} \Delta_{4i}$ and $\sin 2\Delta_{4i}$ average out to 0.5 
and 0 respectively at long baseline ($i = 1,2,3$). 
From Eq.\ \ref{eq:pab} one can use the long baseline approximation (\ie, neglecting the oscillation effects due to $\sdm$) and arrive at the following simplified expression for the 
dominant channel $\nu_{\mu} \to \nu_{e}$.
\begin{align}
\label{eq:pmue}
P_{\mu e}^{4\nu} \approx &\frac{1}{2}\sin^2{2\theta_{\mu e}^{4\nu}} \nonumber \\
&+ (a^2\sin^2 2\theta_{\mu e}^{3\nu} - \frac{1}{4}\sin^2 2\theta_{13}\sin^2{2\theta_{\mu e}^{4\nu}}) \sin^2\Delta_{31} \nonumber\\
&+ \cos(\delta_{13} + \delta_{24}) a \sin{2\theta_{\mu e}^{3\nu}} \sin{2\theta_{\mu e}^{4\nu}}  \cos 2 \theta_{13} \sin^2 \Delta_{31}  \nonumber \\
& + \frac{1}{2} \sin(\delta_{13} + \delta_{24}) a \sin{2\theta^{3\nu}_{\mu e}} \sin{2\theta^{4\nu}_{\mu e}}  \sin{2\Delta_{31}},
\end{align}
where we have followed the convention of \cite{Gandhi:2015xza} for the following quantities.
\begin{center}
    $\sin{2\theta_{\mu e}^{3\nu}} = \sin{2\theta_{13}} \sin{\theta_{23}}$,
    
    $b = \cos{\theta_{13}} \cos{\theta_{23}} \sin{\theta_{12}}$,
    
    $\sin{2\theta_{\mu e}^{4\nu}} = \sin{2\theta_{14}} \sin{\theta_{24}}$,
    
    $a = \cos{\theta_{14}} \cos{\theta_{24}}$.
\end{center}
Eq.\ \ref{eq:pmue} tells us that in vacuum $\pme$ is sensitive to both $\da$ and $\db$, but not to $\dc$. 
As explained in \cite{Gandhi:2015xza}, small dependence on $\dc$ creeps in when matter 
effect is taken into account. 
The expressions for the less dominant channels $\pmm$ and $\pmt$ can similarly be derived from Eq.\ \ref{eq:pab}, 
but the expressions are quite lengthy. 
Interested readers can see \cite{deGouvea:2019ozk, Yue:2019qat} for those probability expressions.  
 \begin{table}[h]
\centering
\scalebox{0.9}{
\begin{tabular}{| c | c | c | c |}
\hline
&&&\\
Parameter & Best-fit-value & 3$\sigma$ interval & $1\sigma$ uncertainty  \\
&&&\\
\hline
&&&\\
$\theta_{12}$ [Deg.]             & 34.3                    &  31.4 - 37.4   &  2.9\% \\
$\theta_{13}$ (NH) [Deg.]    & 8.58              &  8.16  -  8.94   &  1.5\% \\
$\theta_{13}$ (IH) [Deg.]    & 8.63              &  8.21  -  8.99   &  1.5\% \\
$\theta_{23}$ (NH) [Deg.]        & 48.8                     &  41.63  - 51.32    &  3.5\% \\
$\theta_{23}$ (IH) [Deg.]        & 48.8                     &  41.88  - 51.30    &  3.5\% \\
$\sdm$ [$\text{eV}^2$]  & $7.5 \times 10^{-5}$  &  [6.94 - 8.14]$\times 10^{-5}$  &  2.7\% \\
$\ldm$ (NH) [$\text{eV}^2$] & $+2.56 \times 10^{-3}$   &  [2.46 - 2.65] $\times 10^{-3}$ &  1.2\% \\
$\ldm$ (IH) [$\text{eV}^2$] & $-2.46 \times 10^{-3}$  & -[2.37 - 2.55]$\times 10^{-3}$  &  1.2\% \\
$\delta_{13}$ (NH) [Rad.]   & $-0.8\pi$   & $[-\pi, 0]  \cup [0.8\pi, \pi]$ &  $-$ \\
$\delta_{13}$ (IH) [Rad.]   & $-0.46\pi$   & $[-0.86\pi, -0.1\pi]$   & $-$  \\
$\td$ [Deg.] & 5.7, 10 & 0 - 18.4 & $\sigma(\sin^{2}\theta_{14}) = 5\%$ \\
$\te$ [Deg.] & 5, 6 & 0 - 6.05 & $\sigma(\sin^{2}\theta_{24}) = 5\%$ \\
$\tf$ [Deg.] & 20, 25 & 0 - 25.8 & $\sigma(\sin^{2}\theta_{34}) = 5\%$ \\
$\delta_{24}$ [Rad.]   & $0, -0.5\pi$   & $[-\pi, \pi]$ &  $-$ \\
$\delta_{34}$ [Rad.]   & $0, -0.5\pi$   & $[-\pi, \pi]$ &  $-$ \\
&&&\\
\hline
\end{tabular}}
\caption{\label{tab:parameters}
 Standard oscillation parameters and their uncertainties used in our study. The values of 3+0 parameters were taken from the global fit analysis in \cite{deSalas:2020pgw} while the 3+1 parameter values were chosen from \cite{Dentler:2018sju} (see also Sec.\ \ref{sec:basics}).  
If the $3\sigma$ upper and lower limit of a parameter is $x_{u}$ and $x_{l}$ respectively, the $1\sigma$  uncertainty is $(x_{u}-x_{l})/3(x_{u}+x_{l})\%$~\cite{Abi:2020evt}. 
For the active-sterile mixing angles, a conservative $5\%$ uncertainty was used on $\sin^{2}\theta_{i4}$ (i = 1, 2, 3).
}
\end{table}
 \begin{figure}[htb]
 \centering
 \includegraphics[scale=0.36]{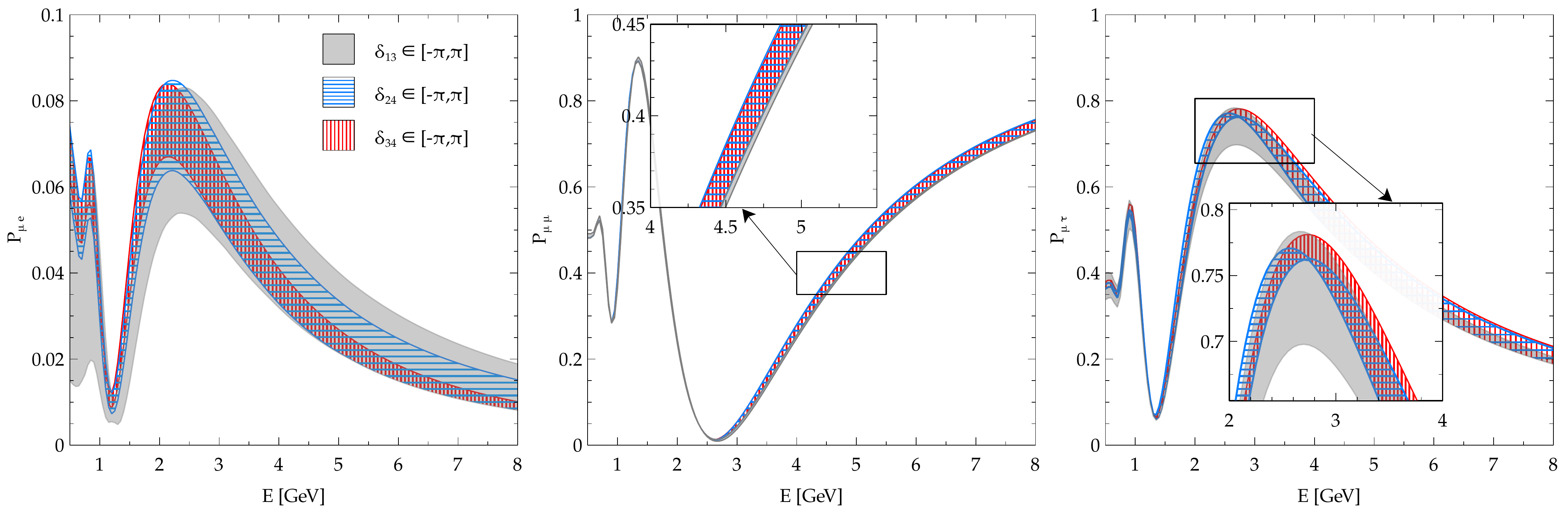}
 \caption{\footnotesize{We show the probability bands due to individual variation of the CP phases 
 $\da$ (grey), $\db$ (blue) and $\dc$ (red) in the whole range of $[-\pi,\pi]$ at a baseline of 1300 km. 
 The three panels correspond to the three channels $\pme, \pmm$ and $\pmt$. 
 The insets in the second and third panels show magnified versions of the rectangular  regions indicated. 
 The three active-sterile mixing angles were taken as $\td = 10^{\circ}, \te = 6^{\circ}, \tf = 25^{\circ}$. 
The values of the rest of the oscillation parameters were taken from Tab.\ \ref{tab:parameters}. Normal hierarchy was assumed for generating this plot. 
 }}
  \label{fig:p_band_d}
 \end{figure}

Using the widely used General Long Baseline Experiment Simulator (GLoBES)~\cite{Huber:2004ka, Huber:2007ji} and the relevant plugin {\it{snu.c}}~\cite{Kopp:2006wp, Kopp:2007ne} for implementing sterile neutrinos, we now illustrate how the probabilities for different oscillation channels depends
on the three CP phases ($\da, \db, \dc$) individually at the DUNE baseline of 1300 km. 
In Fig.\ \ref{fig:p_band_d}, we plot the bands for $P_{\mu e}, P_{\mu\mu}, P_{\mu\tau}$ (in the three 
panels respectively) due to 
individual variation of $\da, \db, \dc$, each being varied in the range $[-\pi, \pi]$.  
The grey band shows the variation of the standard Dirac CP phase $\da$.  
The variation due to $\db$ ($\dc$) is shown with the blue (red) band. 
For the variation of each CP phase, the other two phases are kept fixed: $\da$ is kept 
fixed at the bf value of $-0.8\pi$ while $\db, \dc$ were considered to be zero.  
The sterile phases are associated with the active-sterile mixing angles. 
We have a slightly high active-sterile mixing: $\td = 10^{\circ}, \te = 6^{\circ}, \tf = 25^{\circ}$.
As expected\footnote{In Eq.\ \ref{eq:pab} for $\alpha = \beta$, the imaginary part in the RHS vanishes, diminishing the effect of the CP phases.}, the CP phases have larger impact on appearance channels rather than the disappearance channels. 
$\nue$ channel is most affected by the variation of $\da$. 
Among the sterile CP phases, $\db$ has a visibly greater impact on $\pme$ than that of 
$\dc$. 
It is interesting to note this feature especially in light of the fact that the value of the 
active-sterile mixing angle $\te$ (taken as $6^{\circ}$ in Fig.\ \ref{fig:p_band_d}) is almost 
5 times smaller than $\tf$ ($25^{\circ}$). 
We also observe that with increase in energy the effect of $\dc$ on $\pme$ further reduces.
$\pmt$ is less prone to variation of the CP phases. Still it shows slight variation for all three CP phases 
with $\db$ and $\dc$ having almost similar effects but less than that of $\da$. 
$\pmm$ on the other hand, shows almost negligible variation with the CP phases. 
After a brief description of the simulation procedure we will estimate the capability of LBL experiments to reconstruct these CP phases.

\section{Simulation details}                                                             
\label{sec:analysis}
We simulate the long baseline neutrino experiments DUNE, NOvA, T2K and T2HK using GLoBES~\cite{Huber:2004ka, Huber:2007ji}.
DUNE is a 1300 km long baseline experiment employing a liquid argon far detector (FD) of 40 kt 
fiducial mass with a beam of power 1.07 MW and running 3.5 years each on $\nu$ and 
$\bar{\nu}$ mode (resulting in a total exposure of roughly 300 kt.MW.yr corresponding to 
$1.47 \times 10^{21}$ protons on target or POT). 
We have used the official configuration files~\cite{Alion:2016uaj} provided by the DUNE collaboration for its simulation. 
Following this, we have also taken into account the presence of a near detector (ND) at 459 m from the source. 
The ND helps in making a more precise measurement of the flux and cross-section, thereby reducing 
the relevant systematic uncertainties at the FD. 
We should mention here that we have done a $\chisq$ analysis (discussed later) with the simulated data at FD alone, rather than a joint $\chisq$ analysis using simulated data both at ND and FD\footnote{It is worthwhile to note here that an $\text{eV}$-scale sterile neutrino will have its signature at the ND due to short baseline active-sterile oscillation and consequently a joint analysis using both ND and FD data would probably constrain the active-sterile mixing angles slightly more. 
But our main aim is the analyses of the CP phases, given the already existing constraints on the mixing angles from the global analysis~\cite{Dentler:2018sju}, and this would have more observable signals at the FD, especially in neutrino appearance measurements~\cite{Gandhi:2015xza}}. 
Electron neutrino appearance signals (CC), muon neutrino disappearance signals (CC),  
as well as neutral current (NC) backgrounds and tau neutrino appearance backgrounds (along with the corresponding systematics/efficiencies \etc)
are already included in the configuration files. 

 In the present analysis, we have additionally incorporated tau neutrino appearance 
as a separate signal following \cite{deGouvea:2019ozk, Ghoshal:2019pab}. 
Charged current interaction of an incoming $\nu_{\tau}$ produces a $\tau$ lepton (requires a threshold  energy of $\gtrsim 3.4$ GeV for the incoming $\nu_{\tau}$), which can decay 
hadronically (with a branching fraction $\sim 65\%$) or leptonically (with a branching fraction of $\sim 35\%$). 
The analysis of the hadronic decay channel involves the capability of the detector to study the 
resulting pions and kaons. 
More importantly, NC neutrino scattering constitutes the biggest background for the hadronic decay channel of $\tau$. 
Following  \cite{deGouvea:2019ozk}, we have used an efficiency to separate $30\%$ hadronically decaying $\tau$ events (with about $1\%$ NC events remaining). 
On the other hand, the leptonic decay channels of $\tau$ ($\tau^{-} \to e^{-}\bar{\nu}_{e}\nu_{\tau}$; $\tau^{-} \to \mu^{-}\bar{\nu}_{\mu}\nu_{\tau}$) are more difficult to analyse, due to the large background mainly consisting of $\nu_{e}$-CC and $\nu_{\mu}$-CC respectively (along with backgrounds from NC and contaminations due to wrong sign leptons.). 
Following \cite{Ghoshal:2019pab}, we have taken the efficiency of the electron channel to be $15\%$. 
Due to the overwhelming background, we have taken a nominal efficiency of $5\%$  in the muon channel. 
Naturally, the contribution of the leptonic decay channel of $\tau$ is very small.   
We should also mention here that the decay of $\tau$ at the detector will involve missing energy in the form of an outgoing $\nu_{\tau}$, which in turn makes the energy reconstruction of the incoming $\nu_{\tau}$ difficult. 
From \cite{deGouvea:2019ozk}, we use a Gaussian energy reconstruction with a resolution of $20\%$ which is a conservative estimate. 
We acknowledge that our implementation of the $\nu_{\tau}$ channel as a signal is conservative in nature. Nevertheless, this provides a small but non-negligible statistics in terms of events and $\chisq$ sensitivity (see Sec.\ \ref{sec:channel} and Appendix C). 
Using a much more sophisticated analysis of $\nu_{\tau}$ appearance channel at DUNE by implementing jet-
clustering algorithms and machine learning techniques, as has been pioneered in \cite{Machado:2020yxl}, one certainly expects to exploit the rich physics capabilities hidden within this channel. 


We have simulated NOvA with a baseline of 800 km employing an FD of fiducial mass
of 14 kt and a beam of 742 kW. 
The simulation for NOvA was implemented according to \cite{Acero:2019ksn} which 
generates $8.85 \times 10^{20}$ $(12.33 \times 10^{20})$ POT in $\nu$ ($\bar{\nu}$) 
mode. 
T2K is a 295 km experiment with a 22.5 kt water cherencov FD. 
For T2K simulation we use the inputs from \cite{Abe:2011ks, Abe:2019vii}. 
We have used a beam of 515 kW and simulating $1.97 \times 10^{21}$ $(1.63 \times 10^{21})$ POT in $\nu$ ($\bar{\nu}$) mode. 
T2HK is an {\it{upgraded}} version of T2K with a higher beam of 1.3 MW and a much 
bigger fiducial mass of 187 kt of its water cherencov FD. 
For T2HK we simulate a total of $2.7 \times 10^{22}$ POT in 1:3 ratio of $\nu$ and $\bar{\nu}$ mode (with inputs taken from \cite{Abe:2015zbg, Abe:2018uyc}). 
Note that, for the future experiments DUNE and T2HK we have used the full 
expected exposure, while for the currently running experiments T2K and NOvA we 
have simulated upto their current exposure.   
 

To estimate the capability of LBL experiments to reconstruct the CP phases, we carry out a $\chisq$ analysis 
using GLoBES and the relevant plugin {\it{snu.c}}~\cite{Kopp:2006wp, Kopp:2007ne} for implementing sterile neutrinos. 
In order to gain insight, let us examine the analytical
form of the $\chisq$,
 \begin{align}
\label{eq:chisq}
\Delta \chi^{2}(p^{\text{true}}) = \underset{p^{\text{test}}, \eta}{\text{Min}} \Bigg[&2\sum_{k}^{\text{mode}}\sum_{j}^{\text{channel}}\sum_{i}^{\text{bin}}\Bigg\{
N_{ijk}^{\text{test}}(p^{\text{test}}; \eta) - N_{ijk}^{\text{true}}(p^{\text{true}})\nonumber\\
&+ N_{ijk}^{\text{true}}(p^{\text{true}}) \ln\frac{N_{ijk}^{\text{true}}(p^{\text{true}})}{N_{ijk}^{\text{test}}(p^{\text{test}}; \eta)} \Bigg\} 
+ \sum_{l}\frac{(p^{\text{true}}_{l}-p^{\text{test}}_{l})^{2}}{\sigma_{p_{l}}^{2}}
+ \sum_{m}\frac{\eta_{m}^{2}}{\sigma_{\eta_{m}}^{2}}\Bigg],
\end{align}
where $N^{\text{true}}$ ($N^{\text{test}}$) is the  {\it{true}} ({\it{test}}) set of events, while $p^{\text{true}}$ ($p^{\text{test}}$) 
is the set of {\it{true}} ({\it{test}}) oscillation parameters. 
The index $i$ is summed over the energy bins of the experiment concerned (as discussed above). 
The indices $j$ and $k$ are summed over the oscillation channels ($\nue, \numu, \nutau$) and the modes ($\nu$ and $\bar{\nu}$) respectively.  
The term $(N^{\text{test}} - N^{\text{true}})$ takes into account the algebraic difference while the  log-term inside the curly braces considers the fractional difference  between the {\it{test}} and {\it{true}} sets of events. 
The term summed over $i, j, k$ and written inside the curly braces is the statistical part of $\chisq$.
$\sigma_{p_{l}}$ is the uncertainty in the prior measurement of the $l$-th oscillation parameter $p_{l}$. 
The values of the true or best-fit oscillation parameters and their uncertainties as used in the present analysis are tabulated in Table~\ref{tab:parameters}. 
In the last term, $\sigma_{\eta_{m}}$ is the uncertainty on the systematic/nuisance parameter $\eta_{m}$ and 
the sum over $m$ takes care of the systematic part of $\chisq$. 
This way of treating the systematics in the $\chisq$ calculation is known as the {\it{method of pulls}}~\cite{Huber:2002mx,Fogli:2002pt,GonzalezGarcia:2004wg,Gandhi:2007td}. 
For DUNE, the $\nu_{e}$ and $\bar{\nu}_{e}$ signal modes have a normalization uncertainties of $2\%$ each, whereas the $\nu_{\mu}$ and $\bar{\nu}_{\mu}$ signals have a normalization uncertainty of $5\%$ each. The $\nu_{\tau}$ and $\bar{\nu}_{\tau}$ signals have a normalization uncertainties of $20\%$ each. 
The background normalization uncertainties vary from $5\%-20\%$ and include correlations among various sources of background (coming from beam $\nu_{e}/\bar{\nu}_{e}$ contamination, flavour misidentification, neutral current and $\nu_{\tau}$).  
The final estimate of $\chisq$ obtained after a marginalization over the $3\sigma$ range of test parameters\footnote{The test parameters that are marginalised include those listed in Tab.\ \ref{tab:parameters} along with the specified ranges and priors. Note that, we have used a $5\%$ prior for the sine of each of the active-sterile mixing angles while varying them over their entire range.} $p^{\text{test}}$ and the set of systematics ($\eta$) is a function of the true values of the oscillation parameters. 
Technically this $\chisq$ is the frequentist method 
of hypotheses testing~\cite{Fogli:2002pt, Qian:2012zn}.
 \section{Reconstruction of the CP phase in correlation with other phases}                         
 \label{sec:phase_phase}
 \begin{figure}[htb]
 \centering
 \includegraphics[scale=0.5]{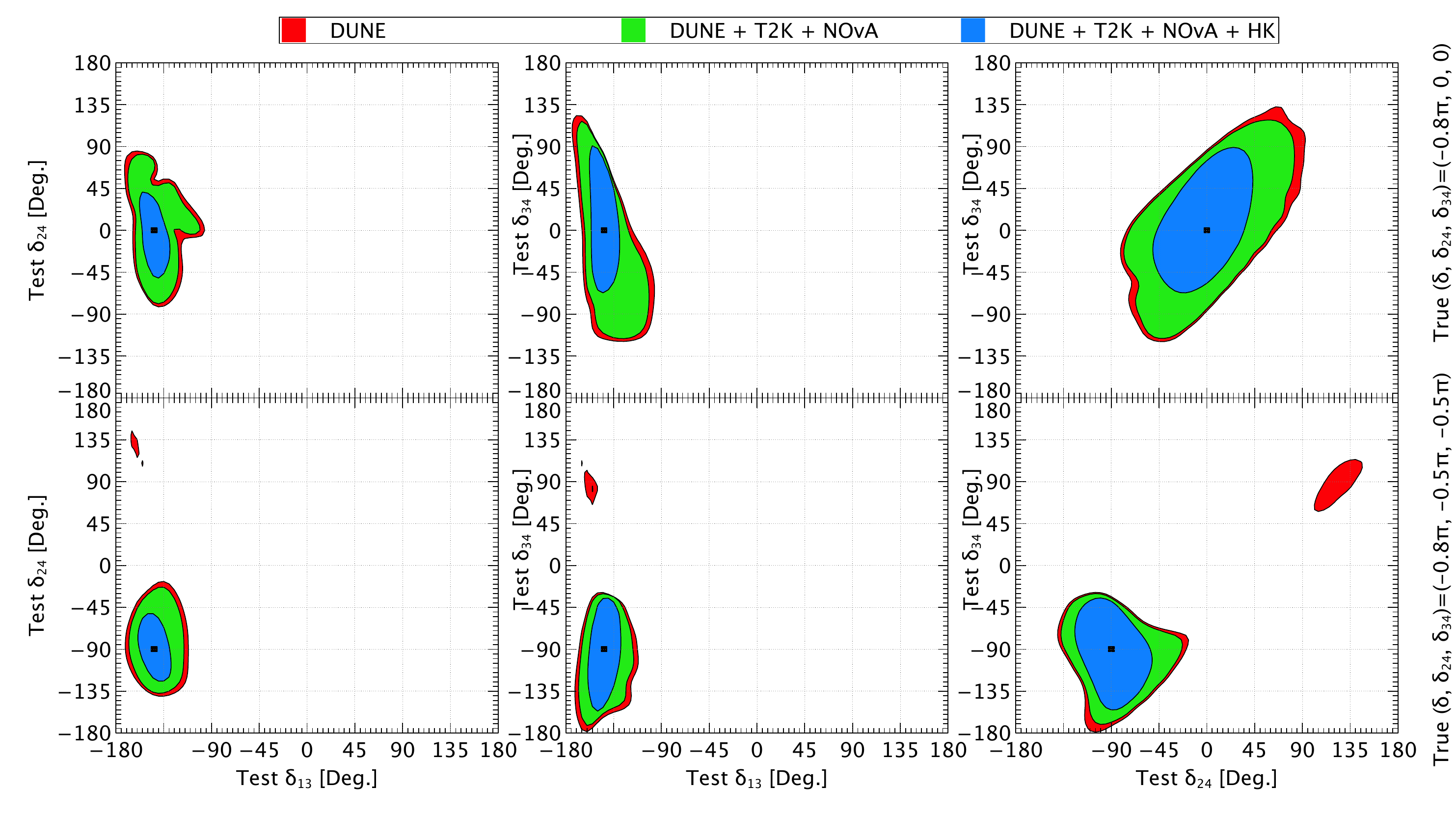}
 \caption{\footnotesize{Reconstruction of the CP phases, taken pairwise at a time, 
 at a C.L. of $1\sigma$ (2 D.O.F.) at DUNE (red), DUNE + T2K + \nova\ (green), and DUNE + T2K + \nova\ + T2HK (blue). 
 The top (bottom) row corresponds to the choice $\db = \dc = 0$ ($-\pi/2$). 
 The true value of the standard Dirac CP phase $\da$ is fixed at $-0.8\pi$. 
 The black dot indicates the true values assumed. 
 The true values of active-sterile mixing angles are taken as $\td, \te, \tf = 5.7^{\circ}, 5^{\circ}, 20^{\circ}$.}
Other oscillation parameters were taken from Tab.\ \ref{tab:parameters}.
 }
  \label{fig:d_d_test}
 \end{figure}
 In Fig.\ \ref{fig:d_d_test}, we illustrate how the CP phases can be reconstructed at $1\sigma$  C.L. in 
 the plane of (test $\da$-test $\db$), (test $\da$-test $\dc$) and (test $\db$-test $\dc$) in the three columns respectively. 
 The red contour shows the reconstruction capability of DUNE. 
 The green and blue contours illustrate the reconstruction when it is combined with
  (T2K + \nova) and (T2K + \nova\ + T2HK) respectively. 
  The top (bottom) row of Fig.\ \ref{fig:d_d_test} shows the choice of the CP conserving (maximally CP violating) {\it{true}} values of $\db$ and $\dc$. 
  Marginalisation has been carried over the test values of $\tb$, $\tc$, $\ldm$ (including both the mass hierarchies) with prior uncertainties mentioned in Tab.\ \ref{tab:parameters}. 
  the active-sterile mixing angles $\td, \te,\tf$ (with a prior uncertainty of $5\%$ on each $\sin^{2}\theta_{i4}$) and the third CP phase not shown along the axes of a particular panel has also been considered for marginalisation.
  Clearly the reconstruction of $\da$ is better  
  (as evidenced by the {\it{narrowness}} of the contours along the test $\da$ axis) 
  compared to the other two phases. 
  This happens since the associated mixing angle $\tb$ has been measured very precisely 
  unlike the corresponding active-sterile mixing angles $\te$ and $\tf$ (see Table \ref{tab:parameters} and Sec.\ \ref{sec:basics}). 
  For the maximally CP violating choices of true $\db$ and true $\dc$, their reconstruction 
  gets better at the cost of small degeneracies appearing for DUNE around test $\db \approx 135^{\circ}$, $\dc \approx 90^{\circ}$. 
  Adding data from other experiments lifts these degeneracies. 
   We would also like to refer the reader to Appendices A and B, which show the effect of  
 prior of $\tc$ and different {\it{true}} choices of the standard oscillation parameters on the reconstruction of the CP phases. 
   \begin{figure}[htb]
 \centering
 \includegraphics[scale=0.5]{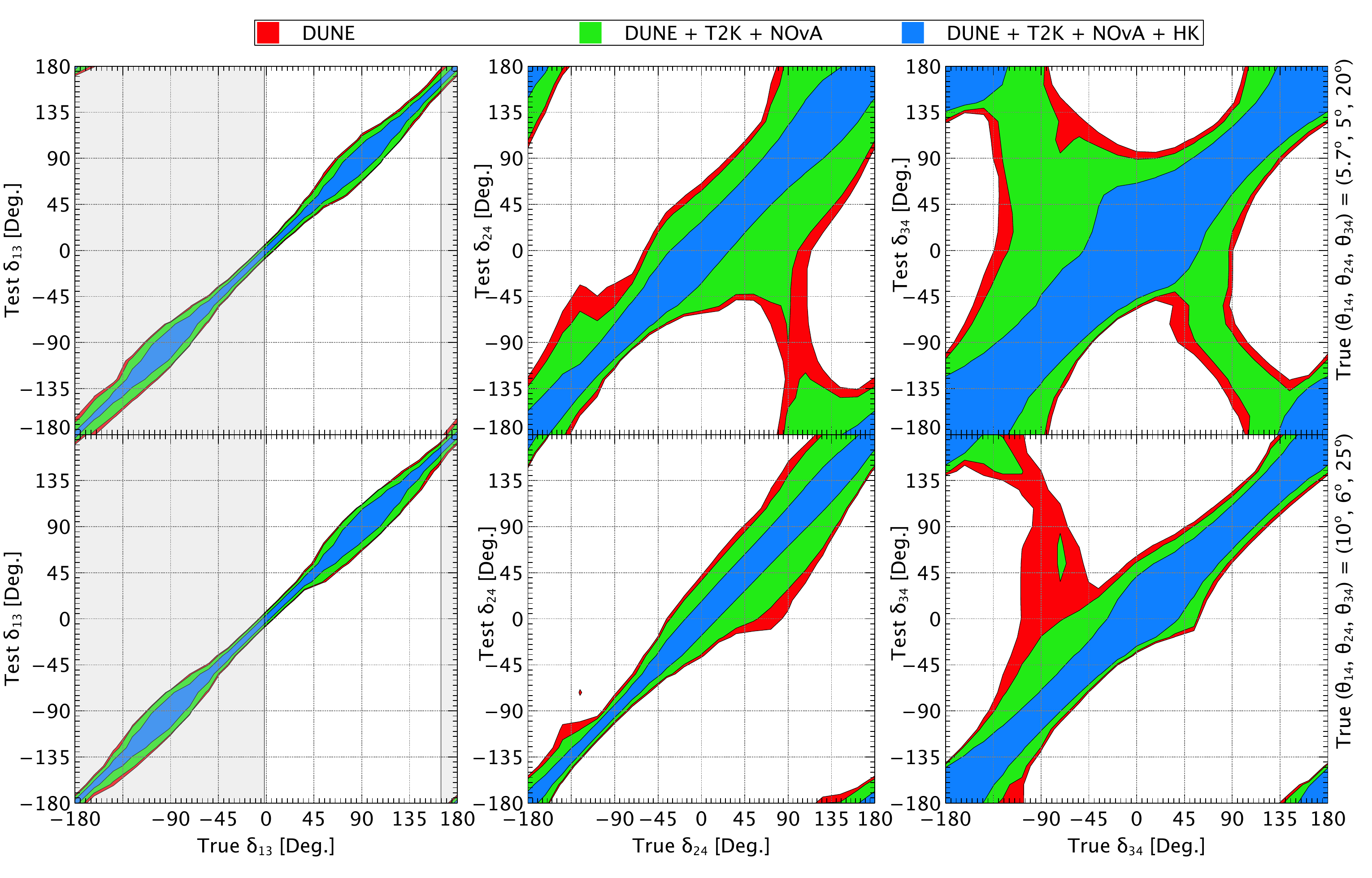}
 \caption{\footnotesize{Reconstruction of the CP phases $\da$, $\db$ and $\dc$, for all the choices of their true values in $[-\pi,\pi]$
 at a C.L. of $1\sigma$ (1 D.O.F.) at DUNE (red), DUNE + T2K + \nova\ (green), and DUNE + T2K + \nova\ + T2HK (blue). 
 The top (bottom) row corresponds to the choice of the active-sterile mixing angles as true $\td, \te, \tf = 5.7^{\circ}, 5^{\circ}, 20^{\circ}$ ($10^{\circ}, 6^{\circ}, 25^{\circ}$).  
 The true values of the phases not shown in a panel are fixed at: $\da$, $\db$ and $\dc = -0.8\pi, 0$ and $0$ respectively.
 In the first column the grey shaded regions depict the $3\sigma$ allowed values measured by T2K~\cite{Abe:2019vii}.
 }}
  \label{fig:d_d_true}
 \end{figure}
 
  In Fig.\ \ref{fig:d_d_true}, we illustrate how efficiently the combination of LBL experiments can reconstruct the three CP phases $\da, \db$ and $\dc$ at $1\sigma$ C.L. in the three columns respectively given their true value lying anywhere in the whole parameter space of $[-\pi, \pi]$. 
   In addition to the poorly measured 3+0 parameters ($\tc, \ldm$) and the active-sterile mixing angles ($\td, \te, \tf$), in each panel we have also marginalised over the two other CP phases ($\in [-\pi, \pi]$) not shown along the axes.  
  The top (bottom) row depicts small (large) 
  active-sterile mixing with true $\td, \te, \tf = 5.7^{\circ}, 5^{\circ}, 20^{\circ}$ ($\td, \te, \tf = 10^{\circ}, 6^{\circ}, 25^{\circ}$). 
  Note that, the latter choice of values represents the upper limits of the allowed active-sterile mixing (See Sec. \ref{sec:basics}).
  For each true value of the CP phases ($\in [-\pi, \pi]$), the corresponding vertical 
  width of the contours provide an estimate of the precision of reconstructing that 
  true value. 
  It is evident that in comparison to the reconstruction of the sterile phases, the standard Dirac phase $\da$ can be reconstructed much more efficiently by the LBL experiments.  
  We note that the reconstruction of $\da$ does not noticeably depend upon 
  the size of the active-sterile mixing as assumed in the two rows.  
  As the T2K data~\cite{Abe:2019vii} suggests, if $\da$ indeed turns out to lie in 
  the lower half plane of $[-\pi, 0]$ with a best fit roughly around the maximal CPV ($\approx -\pi/2$), the future analyses 
  with a combination of (DUNE + T2K + \nova + T2HK) with their projected runtime 
 will be able to measure this phase in a narrow approximate range of $[-115^{\circ}, -75^{\circ}]$ (at $1\sigma$). 
 The precision is marginally better if $\da$ turns out to be close to the CP conserving value ($0$).
 As far as the reconstruction of the sterile phases are concerned, $\db$ can be 
 reconstructed with a better precision than $\dc$. 
 For small active-sterile mixing (top row of Fig.\ \ref{fig:d_d_true}) and without considering the T2HK-projected data, $\db$ can be better reconstructed only in the lower half plane. 
 But with T2HK-projected data, its reconstruction becomes much better throughout 
 the entire parameter space. 
 T2HK, due to its shorter baseline and much higher fiducial mass of its water cerenkov detector can offer very high statistics which helps to alleviate the degeneracy. 
 On the other hand, the reconstruction of $\dc$ is bad almost for the entire parameter space if T2HK is not considered. 
 For \eg, if the true value of $\dc$ in nature turns out to be around $[-135^{\circ}, -90^{\circ}]$, the range of 
  values reconstructed at $1\sigma$ by the combination of DUNE + T2K + \nova\ 
  can be anywhere between $-180^{\circ}$ to $180^{\circ}$.  
  Large active-sterile mixing (bottom row of Fig.\ \ref{fig:d_d_true}) significantly improves the sensitivities of the LBL 
  experiments to the sterile CP phases, resulting in significant improvement of the 
  reconstruction of $\db$ and $\dc$. 
  We have already seen in Fig.\ \ref{fig:p_band_d} that $\pme$ is most sensitive to $\da$ and least sensitive to $\dc$.
  This leads to a better precision in reconstructing $\da$ in comparison to the others.

  \section{Role of individual channels in reconstruction}                                       
 \label{sec:channel}
\begin{figure}[b]
 \centering
 \includegraphics[scale=0.5]{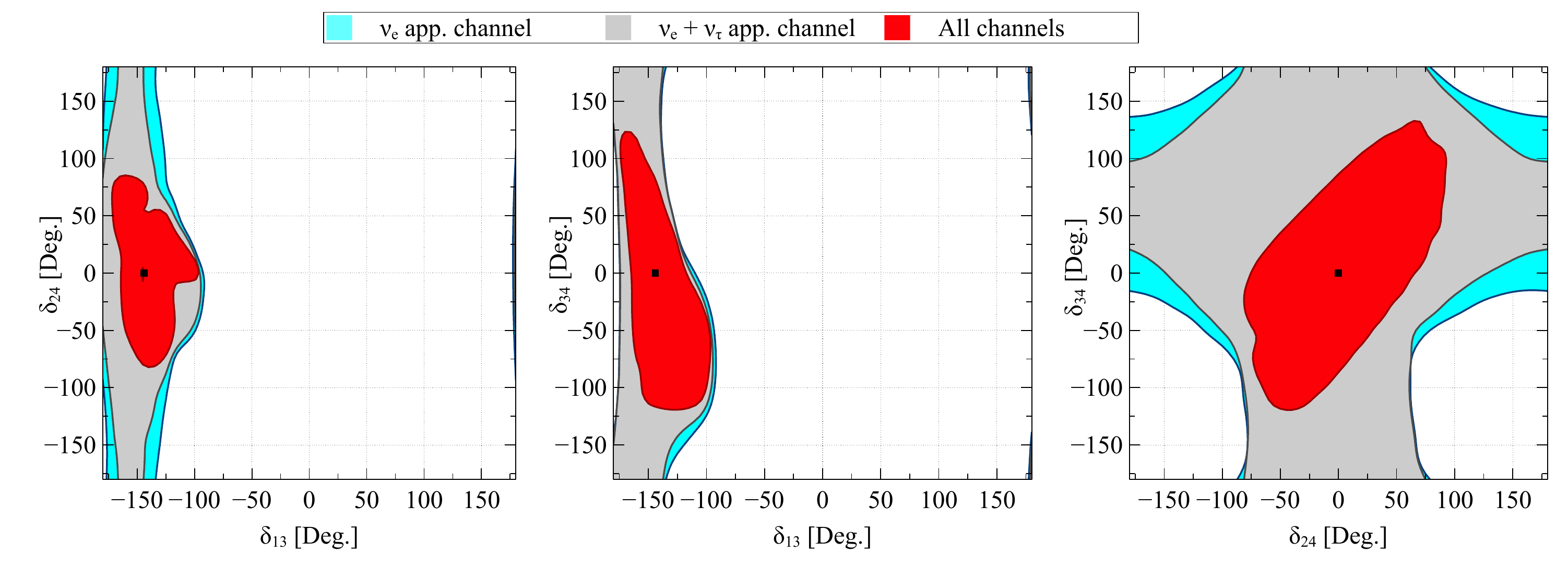}
 \caption{\footnotesize{Reconstruction of the CP phases, taken pairwise at a time for three different channels at DUNE at a C.L. of $1\sigma$ (2 D.O.F.). 
 for $\nu_e$ appearance channel (cyan), $\nu_e$ +  $\nu_{\tau}$ appearance channel (grey) and all channels (red). 
 The three panels depict the test values of the three CP phases. 
 The true values for the CP phases were assumed as $\da, \db, \dc = -0.8\pi, 0, 0$ and the true active-sterile mixing angles were chosen as $\td, \te, \tf = 5.7^{\circ}, 5^{\circ}, 20^{\circ}$. 
 The black dot indicates the true values assumed.
 }}
  \label{fig:d_d_channel}
 \end{figure}

 Fig.\ \ref{fig:d_d_channel} illustrates the impact of different appearance and disappearance channels on the reconstruction of the CP phases (taken pairwise, like in Fig.\ \ref{fig:d_d_test}) at $1\sigma$ C.L. at DUNE.  
    The innermost red contours consisting of all the three channels for DUNE are the same as the red contours in the upper row of Fig.\ \ref{fig:d_d_test}. 
  We can clearly see the decrease in uncertainty in the measurement of the CP phases as we keep on adding $\nu_{\mu} \rightarrow \nu_{\tau}$ appearance and $\nu_{\mu} \rightarrow \nu_{\mu}$ disappearance channel to the $\nu_{\mu} \rightarrow \nu_{e}$ appearance channel.
 The phase dependence of the $\nutau$ channel (see also Fig.\ \ref{fig:p_band_d}) helps somewhat in this case. 
  Note that, the improvement in result due to the addition of the $\nu_{\tau}$ appearance channel is not very significant and this is due to the very low statistics provided by this channel due to the difficulty in observing $\nu_{\tau}$. 
  On the other hand, though the phase dependence of $\numu$ channel is small, it offers a very large number of events compared to the other two channels, thereby decreasing the uncertainty in reconstruction (see also, Appendix C).
  The improvement in reconstruction along the $\da$ direction is not significant with the addition of channels, underlining the pivotal role here played by the $\nu_{e}$ appearance channel. 
  In contrast, as observed from the first two panels of Fig.\ \ref{fig:d_d_channel}, the 
  reconstruction of $\db$ 
  and $\dc$ is significantly better (more so in the former) particularly by the addition of $\nu_{\mu}$ disappearance channel.
 In the third panel, the role of $\nu_{\mu}$ disappearance is emphasized by the 
 shrinking of the reconstruction contour in both $\db$ and $\dc$ directions. 
 Here we also note a slight improvement in reconstruction along the $\dc$ axis (but not so much along $\db$) with the addition of $\nu_{\tau}$ appearance channel. 
 In the next section we are going to estimate the correlations of the phases with the active-sterile mixing angles.
 \section{Reconstruction of the CP phase in correlation with active-sterile mixing angles}                 
 \label{sec:phase_angle}
\begin{figure}[htb]
 \centering
 \includegraphics[scale=0.52]{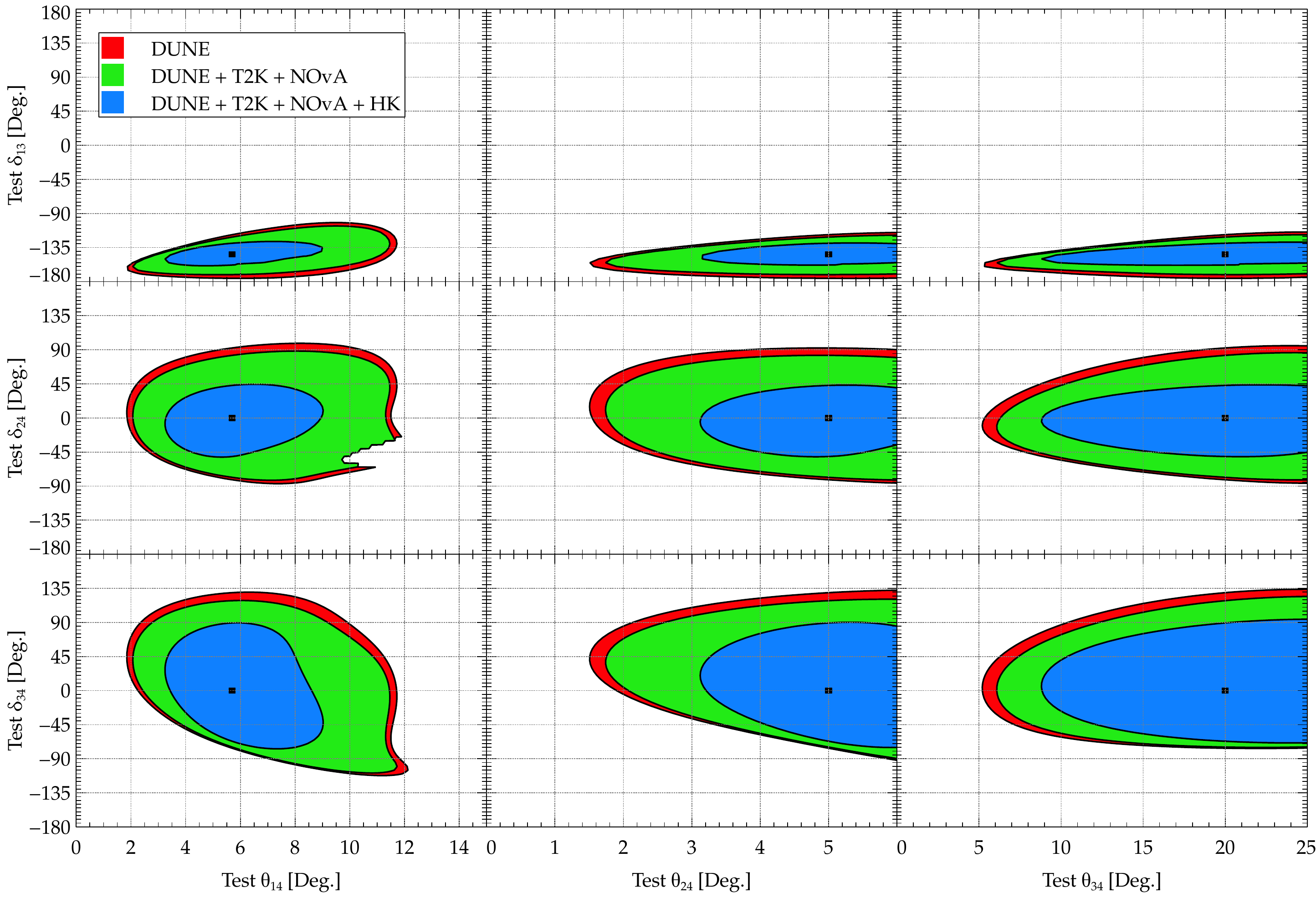}
 \caption{\footnotesize{Reconstruction of the CP phases in correlation with the mixing angles $\td, \te, \tf$
 at a C.L. of $1\sigma$ (2 D.O.F.) at DUNE (red), DUNE + T2K + \nova\ (green), and DUNE + T2K + \nova\ + T2HK (blue). 
 The three columns (rows) depict the test values of the three active-sterile mixing angles (three CP phases). 
 The true values for the CP phases were assumed as $\da, \db, \dc = -0.8\pi, 0, 0$ and the true active-sterile mixing angles were chosen as $\td, \te, \tf = 5.7^{\circ}, 5^{\circ}, 20^{\circ}$. 
 The black dot indicates the true values assumed.
 }}
  \label{fig:d_th_test}
 \end{figure}

 In Fig.\ \ref{fig:d_th_test} we show how efficiently the LBL experiments can reconstruct the CP phases 
 in correlation with active-sterile mixing angles at $1\sigma$ C.L. (2 D.O.F.). 
 The three phases (angles) are shown along the three rows (columns). 
In the second and third column the ranges of $\theta_{24}$ and $\theta_{34}$ are not shown beyond their allowed values.
This results in the contours not being closed in these cases. 
As expected, the top row shows much less uncertainty along the test $\da$ direction for all the three angles involved. 
In all the panels, we note that combining \nova\ and T2K data to DUNE gives mild improvement while a further addition of data projected by T2HK significantly improves the reconstruction. 
We have varied the test values of the angles upto their upper limit at $99\%$ C.L.: 
$\td \lesssim 18.4^{\circ}, \te \lesssim 6.05^{\circ}, \tf \lesssim 25.8^{\circ}$, and the 
true values to be reconstructed were $\td = 5.7^{\circ}, \te = 5^{\circ}, \tf = 20^{\circ}$ 
(as discussed in Sec.\ \ref{sec:basics}). 

The horizontal (vertical) span of the resulting contours indicate the uncertainties 
along the mixing angles (CP phases). 
We observe that the contours involving $\te$ and $\tf$ do not close as long as we confine 
ourselves upto the allowed upper limit of $\te$ and $\tf$. 
It is interesting to note that even if $\tf$ is allowed to vary in a larger range, the contours 
involving it show considerable uncertainty. 
This signifies the difficulty in constraining $\tf$ in the experiments. 
It is mainly this reason which also hinders a good reconstruction of the associated phase 
$\dc$, as we have seen in Sec.\ \ref{sec:phase_phase}. 
To have a quantitative idea about the potential to reconstruct the true values of the active-sterile mixing angle $\theta_{i4}$ ($i = 1, 2, 3$), we calculate how much the total horizontal span of each blue contour is. 
From this we estimate the maximum range of uncertainty (so as to obtain a conservative range) in reconstructing $\theta_{i4}$ and tabulate these below in Table \ref{tab:theta_recon}.    
 \begin{table}[htb]
\begin{center}
\begin{tabular}{|c|c|c|}
\hline
Angle & Value to be reconstructed & Reconstructed range \\
& [Degree] & [Degree]\\
\hline
$\td$ & 5.7 & $3.3 \lesssim \td \lesssim 9.1$\\ \hline
$\te$ & 5 & $3.2 \lesssim \te \lesssim 6$\\ \hline
$\tf$ & 20 & $9.0 \lesssim \tf \lesssim 25$\\ \hline
\end{tabular}
\end{center}
\caption{The maximum reconstructed ranges for $\td, \te, \tf$ as estimated from 
Fig.\ \ref{fig:d_th_test} for DUNE + T2K + NOvA + HK.}
\label{tab:theta_recon}
\end{table}%
  \section{Impact of constraints on Neutrinoless Double Beta Decay}
  \label{sec:ndbd}
Before concluding, we include discussion of $3+1$ framework on other observables, such as, neutrinoless double beta decay. The sterile neutrino, if a  Majorana particle, gives non-zero contribution in the lepton number violating neutrinoless double beta decay (NDBD).  In the presence of a non-zero $\theta_{14}$, the effective mass of NDBD process becomes\cite{Giunti:2015kza}, 
 \begin{eqnarray}
 m_\textrm{eff}=|m_1 |U_{e1}|^{2}+m_2 |U_{e2}|^{2} e^{i \alpha_{2}}+ m_3 |U_{e3}|^{2} e^{i \alpha_{3}}+m_4 |U_{e4}|^{2}e^{i \alpha_{4}}|,
 \label{eq:mefff}
  \end{eqnarray}
  where $\alpha_{2}$, $\alpha_{3}$, $\alpha_{4}$ are the relevant CP phases.
  Following the parametrisation given in Eq.~\ref{eq:param}, the relevant elements of the mixing matrix are as follows. 
  \begin{equation}
  |U_{e1}|=c_{12}c_{13}c_{14}, |U_{e2}|=s_{12}c_{13}c_{14}, |U_{e3}|=s_{13} c_{14}, |U_{e4}|=\sin \theta_{14}.
\label{e:uei}
\end{equation}

 {The expression for half-life of $0\nu\beta\beta$ transition can be given as~\cite{Mitra:2011qr},
\begin{equation}
\label{e:Thalf1}
\frac{1}{T_{1/2}^{0\nu}} = G_{0\nu}\left| M_{\nu} \eta_{\nu} + M_{N} \eta_{N} \right|^2,
\end{equation}
where,
\begin{equation}
\eta_{\nu} = \frac{U_{e\;i}^{2}m_{i}}{m_{e}}, ~~~ \eta_{N} = \frac{V_{e\;i}^{2}m_{p}}{M_{i}}.
\end{equation}}

In the above, $m_{i}$ is the mass of active neutrino and $U_{ei}$ is the PMNS mixing; whereas, $\theta_{ei}$ is the mixing among the active and sterile and $M_{i}$ is the corresponding mass of heavy sterile. In the above,  $M_{\nu}$ and $M_{N}$ are the nuclear matrix elements (NME) for exchange of light and heavy neutrinos respectively. 
   \begin{figure}[htb]
 \centering
  \includegraphics[scale=0.67]{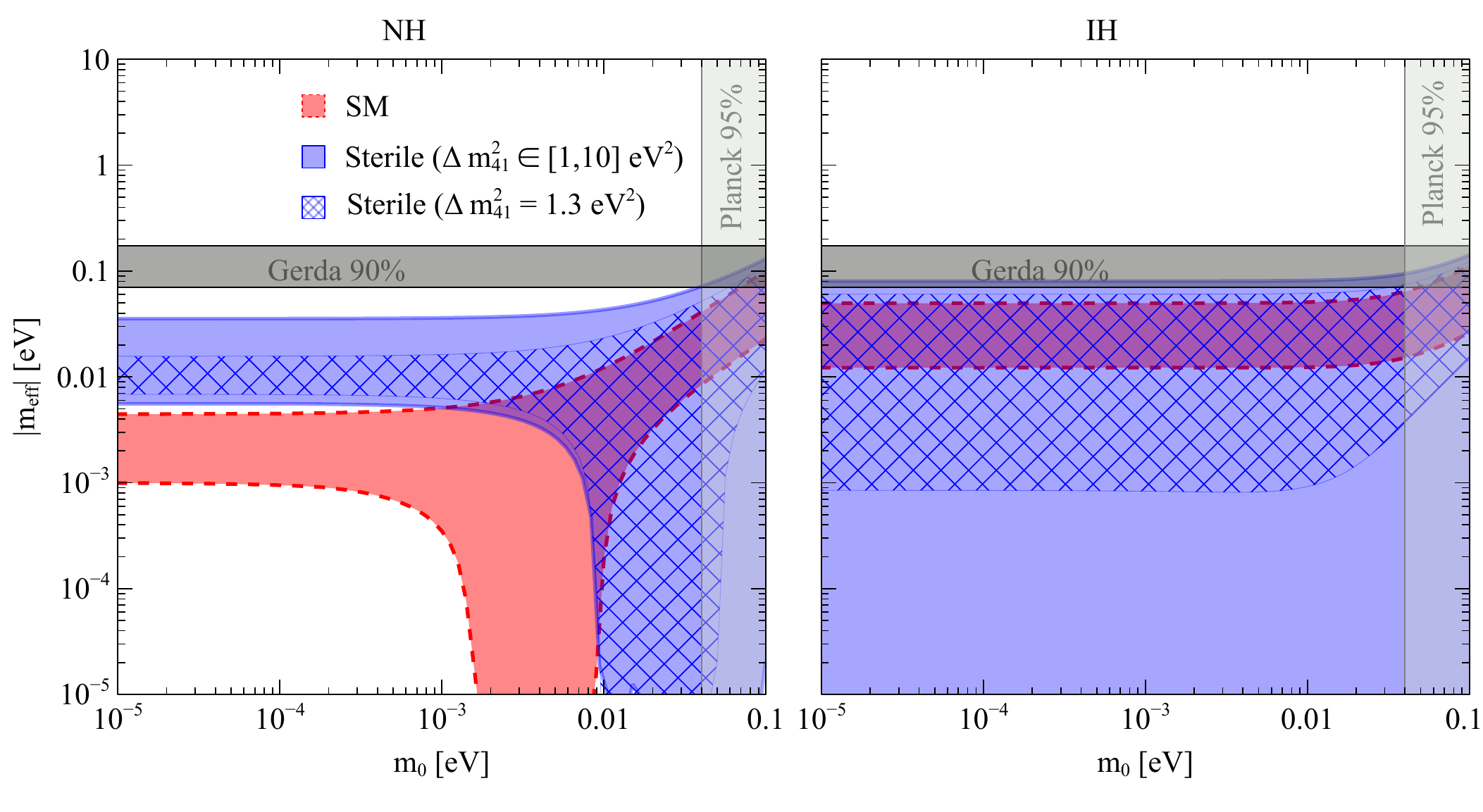}
 \caption{\footnotesize{Effective mass of NDBD versus the smallest neutrino mass. 
 The left (right) panel shows the case of NH (IH). The red (blue) region correspond to 
  3+0 (3+1) scenario, labelled as SM (Sterile). 
  The region hatched with blue {\it{cross}} lines represents $\lldm = 1.3 \text{ eV}^{2}$.
  The vertical light grey region is excluded by Planck data at $95\%$ C.L.~\cite{Aghanim:2018eyx}, while the horizontal dark region shows the $90\%$ sensitivity from GERDA~\cite{Agostini:2019hzm}.}}
    \label{fig:meff_m}
 \end{figure}
The values of NME and phase space factor $G_{0\nu}$ can be find in  Ref.~\cite{Meroni:2012qf}.  The  half-life of $0\nu 2\beta$ is can generally given as~\cite{Kovalenko:2009td} 
\begin{equation}
\label{e:Thalf2}
\frac{1}{T_{1/2}}=K_{0\nu} \left| \Theta_{ej}^2 \frac{ \mu_j}{\langle p^{2} \rangle-\mu^{2}_{j}}  \right|^{2},
\end{equation}
where $j$ represents the number of light neutrino states and the additional heavy neutrino states.  The parameters $\mu_{j}$ and $\Theta_{ej}$  represent the masses of the neutrino states and the mixing with SM neutrinos respectively. In the above,  $K_{0\nu} = G_{0\nu} (\mathcal{M}_{N} m_{p})^{2}$ and  $\langle p^{2} \rangle\equiv -m_{e} m_{p} \frac{\mathcal{M}_{N}}{\mathcal{M}_{\nu}}$.
Over the decade, many experiments on improving the lower limits on $T_{1/2}^{0\nu}$ for $0\nu\beta\beta$ transition have been performed and best stringent bounds have been acquired from germanium-76, Xenon-136 and tellurium-130~\cite{Alduino:2017ehq} isotopes. The experiment GERDA-II, at $90\%$ C.L. has obtained the corresponding lower limit as $T_{1/2}^{0\nu} > 8.0 \times 10^{25}$ year for $Ge^{(76)}$~\cite{Agostini:2018tnm}, whereas from KamLAND-Zen experiment $Xe^{(136)}$, at $90\%$ C.L. the lower limit on the half-life is obtained as $T_{1/2}^{0\nu} > 1.07 \times 10^{26}$ year~\cite{KamLAND-Zen:2016pfg, Penedo:2018kpc}.

In presence of a sterile neutrino, the effective mass obtained from $3+0$ framework will be significantly changed. Using Eq.\ \ref{eq:mefff}, in Fig.~\ref{fig:meff_m}, we show the effective mass for the standard $3+0$ scenario, and for $3+1$ scenario with the variation of the lightest mass. The left and right panels represent NH and IH, respectively. 
 We consider a variation of $\Delta m^2_{41}$ (for NH), and $\Delta m^2_{43}$ (for IH) in between (1-10) $\textrm{eV}^2$.
 The CP phases $\alpha_{2}$, $\alpha_{3}$ and $\alpha_{4}$ have been varied in between $-\pi$ to $\pi$. 
 Other oscillation parameters 
 have been varied in their $3\sigma$ ranges as shown in Tab.\ \ref{tab:parameters}~\cite{deSalas:2020pgw} as applicable for NH and IH. 
Overall we have considered $10^{7}$ iterations at each value of the lightest mass in generating Fig.\ \ref{fig:meff_m}.   
 The red and blue regions represent the variation of $|m_{eff}|$ for the standard $3+0$ scenario, and $3+1$ scenario respectively.  As can be seen from the figure the $|m_{eff}|$ can be significantly large in the presence of a sterile neutrino with large mass value, and hence constrained from the experimental constraint. 
 To make a connection with our oscillation analyses in the preceding sections, we shade 
 the regions corresponding to $\lldm = 1.3 \text{ eV}^{2}$ with blue {\it{cross}} lines. 
 Consequently we note a slight shrinking of the blue regions and it largely comes down to 
 below the exclusion limit by Gerda. 
We note that for NH, there is a complete cancellation of $m_\textrm{eff}$ in the 3+0 case when the lightest mass  approximately lies in the range of $10^{-3}-10^{-2} \text{ eV}$. 
In 3+1 case, there is no complete cancellation in this region (due to the dominance of the $m_{4}$-term in Eq.\ \ref{eq:mefff} in this range) and it only happens when the lightest mass becomes $10^{-2}$ eV and beyond. 
For IH, the standard 3+0 case shows no total cancellation (dominant $m_{1}$-term in Eq.\ \ref{eq:mefff}), while there is a total cancellation in the 3+1 case in the range shown. 
Our results qualitatively agree with \cite{Giunti:2015kza}.

 \section{Summary and Conclusion}
 \label{sec:summary}
 In this paper we have considered the presence of an eV-scale sterile neutrino (the 
 so called 3+1 scenario which might turn out to be a possible resolution of the 
 short baseline neutrino oscillation anomalies) and have analyzed how the present and 
 future long baseline experiments T2K, NOvA, DUNE and T2HK can potentially probe the additional CP phases. 
 We discuss how the three CP phases, namely $\da, \db$ and $\dc$ can individually 
 affect the oscillation channels under consideration and appear in the probability expression. 
 In light of the constraints on the active-sterile mixing from the global analysis, we 
 estimate how the LBL experiments can probe the parameter spaces associated to 
 the CP phases, by taking a pair of CP phases at a time. 
 Though $\nue$ oscillation channel contributes the most in probing these parameter 
 spaces, $\numu$ and to a lesser extent $\nutau$ channel also help in exploring the 
 $\db-\dc$ parameter space in particularly.
 By marginalizing over all other parameters we then show how the three individual CP phases can be reconstructed for all 
 possible true values in the whole range of $[-\pi, \pi]$. 
 We find that $\db$ and $\dc$ cannot be reconstructed very efficiently by DUNE and also 
 even after adding data from NOvA and T2K. 
 But adding T2HK data removes much of the degeneracies and the uncertainties in  reconstruction become much less. 
 We found that if the active-sterile mixing angles turn out to be lying close to their 
 current upper limits, the enhanced sensitivities to the associated phases make the 
 reconstructions of $\db$ and $\dc$ much better. 
 In contrast the reconstruction of the standard CP phase $\da$ is much better even in 
 presence of a light sterile neutrino and this conclusion is almost independent of the 
 size of active-sterile mixing. 
We then analyze how efficiently the experiments can probe all the parameter spaces 
associated to one CP phase and one active-sterile mixing angle. 
It turns out that the parameter regions connected to the angle $\td$ can be probed 
relatively better that those related to the other two mixing angles. 
Finally, we briefly show how the relevant parameter spaces in $0\nu\beta\beta$ get 
modified in light of the active-sterile constraints used in this analysis.  
 
 \acknowledgments
N.F. is grateful for a visit to IOP, Bhubaneswar where this project was initialized. 
N.F. acknowledges the support of Dr.\ Satyajit Jena of IISER Mohali. 
M. Masud acknowledges Dr. S.K. Agarwalla of IOP, Bhubaneswar for providing the financial support from the Indian National Science Academy (INSA) Young Scientist Project [INSA/SP/YS/2019/269]. 
M. Masud is supported by IBS under the project code IBS-R018-D1. 
M. Mitra acknowledges the financial support from DST INSPIRE Faculty research grant (IFA-14-PH-99), and thanks Indo-French Centre for the Promotion of Advanced Research for the support (grant no: 6304-2).
We are grateful to the anonymous referee for constructive suggestions and inputs.

\appendix
\renewcommand{\theequation}{C \arabic{equation}}
\setcounter{equation}{0} 
\section*{Appendix A: Effect of $\tc$-uncertainty}
\label{apndx_a} 
 \begin{figure}[htb]
 \centering
 \includegraphics[scale=0.47]{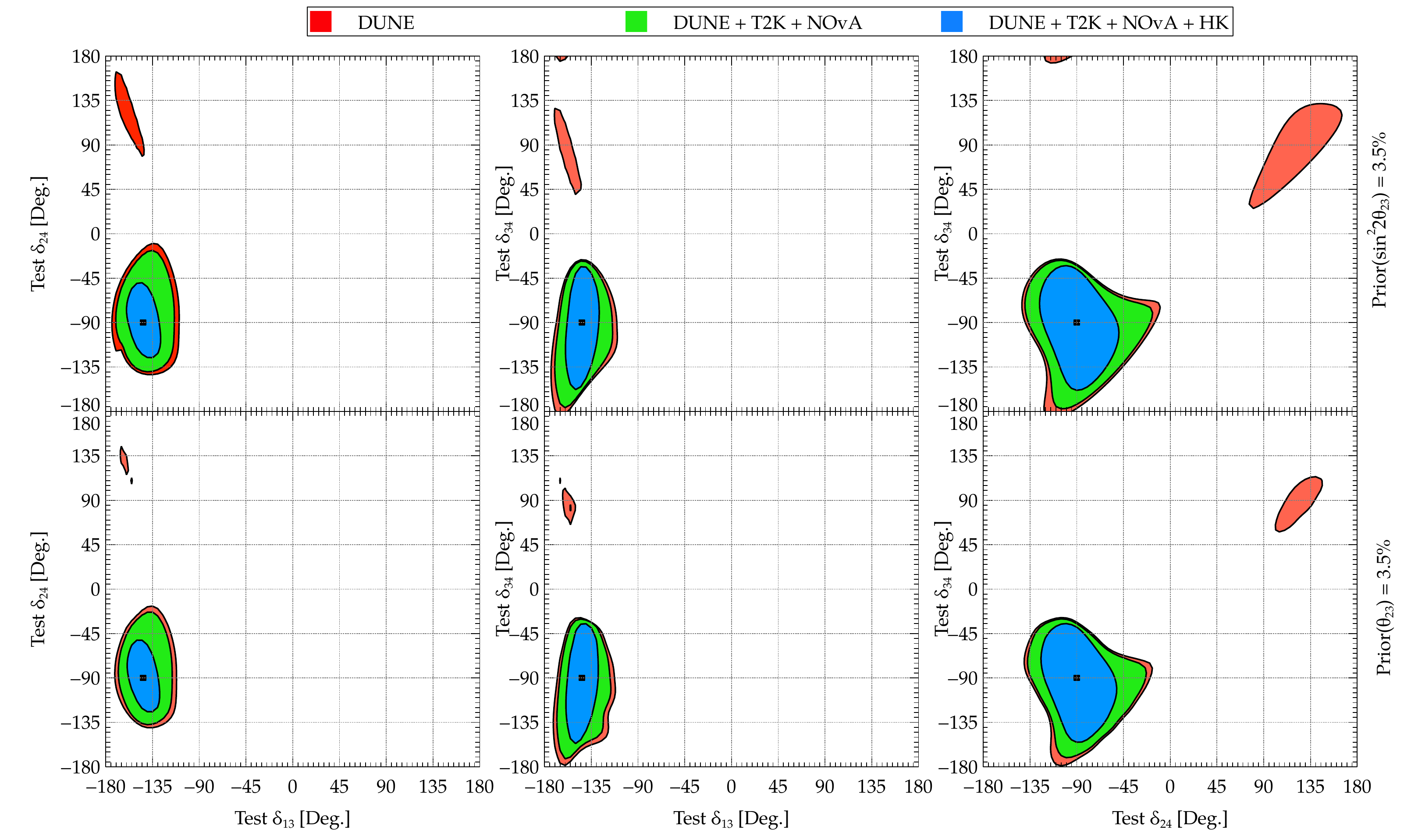}
 \caption{\footnotesize{The effect of taking a non-Gaussian prior uncertainty on $\tc$ on the reconstruction of the CP phases, taken pairwise at a time, 
 at a C.L. of $1\sigma$ (2 D.O.F.) for DUNE (red), DUNE + T2K + \nova\ (green), and DUNE + T2K + \nova\ + T2HK (blue). 
 The black dots represent the true values to be reconstructed. 
 The top row shows the case of taking a prior uncertainty of $3.5\%$ on $\sin^{2}2\tc$. 
 The bottom row (same as that of the bottom row of Fig.\ \ref{fig:d_d_test}), depicts the case of a 
 Gaussian prior uncertainty of $3.5\%$ on $\tc$. 
 }}
  \label{fig:d_d_test_th23_prior}
 \end{figure}
 In all the results so far, we have used a Gaussian prior of $3.5\%$ on $\tc$. 
 Since a true value lying in the higher octant of $\tc$ has been used in the analyses, we also need 
 to study the effect of using a non-Gaussian prior in order to properly consider both the octants when 
 we marginalise over $\tc$. 
 For this purpose, we take a prior uncertainty of $3.5\%$ on $\sin^{2}2\theta_{23}$ in the reconstruction 
 of the CP phases and compare it with the case of considering a Gaussian prior in Fig.\ \ref{fig:d_d_test_th23_prior}.   
We note that, taking the prior on $\sin^{2}2\tc$ results in only a mild spread in the contours, especially  along the axes of the sterile CP phases $\db$ or $\dc$. 
But this does not change our overall results qualitatively.
For \eg, though the contours showing the degenerate solutions at DUNE (red) gets bigger with a prior on $\sin^{2}2\tc$, addition of data from other experiments still removes this degenerate fake solution as before. 
\section*{Appendix B: Effect of different {\it{true}} choice of the standard parameters}
\label{apndx_b}
\begin{figure}[htb]
 \centering
 \includegraphics[scale=0.47]{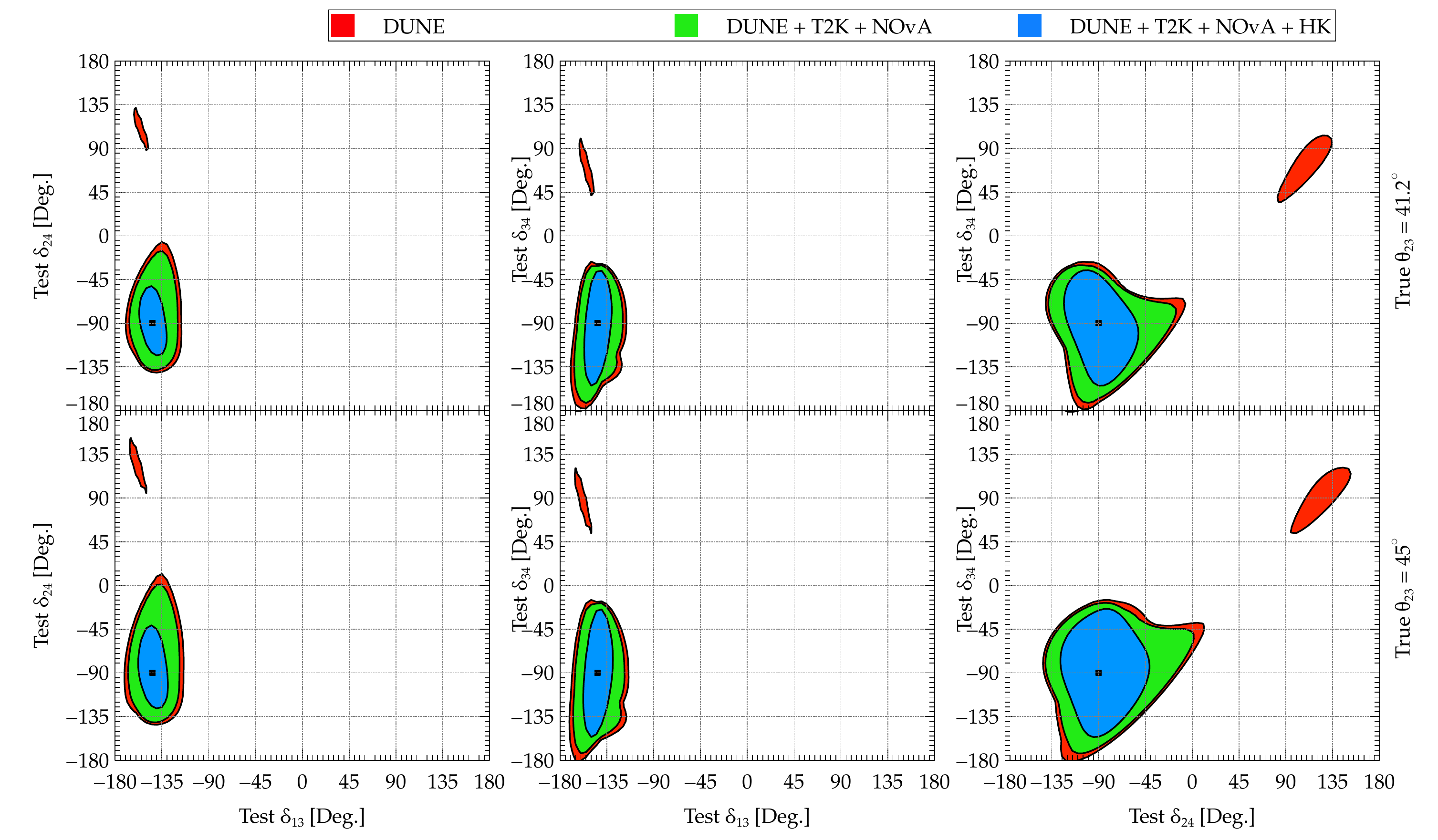}
 \caption{\footnotesize{Effect of considering different {\it{true}} $\tc$-octant while probing the CP phases, taken pairwise at a time (same as the bottom row of Fig.\ \ref{fig:d_d_test_th23_prior}, but for lower octant and maximal mixing for $\tc$). 
 }}
  \label{fig:d_d_test_octant}
 \end{figure}
The best fit values of the standard oscillation parameters used in our 
analyses (as mentioned in Tab.\ \ref{tab:parameters}) were obtained under the 3+0 scenario. 
It has been shown in literature that some of these standard parameters are prone to significant changes 
in the presence of a light sterile neutrino. 
For \eg, the values of $\tc$, $\ldm$, $\da$ can change so much that the issues of correct octant, mass 
hierarchy or CP Violation become very ambiguous (see for \eg, \cite{Agarwalla:2016xlg, Dutta:2016glq}). 
The reactor mixing angle $\tb$, on the other hand is quite robust even in the 3+1 scenario~\cite{Kopp:2013vaa}. 
Though the marginalisation process during $\chisq$ calculation partially takes care of the uncertainties 
of relevant parameters in the {\it{fit}}, a brief discussion regarding the effect of different {\it{true}} values of the standard parameters are in order. 
In Fig.\ \ref{fig:d_d_true}, we had already illustrated our results for all possible true values of the CP violating phases. 
Here we now show in Fig.\ \ref{fig:d_d_test_octant}, the impact of the {\it{true}} choice of lower octant (top row) and maximal mixing (bottom row) of $\tc$ in probing the CP phases. 
 This is similar to the bottom row of Fig.\ \ref{fig:d_d_test_th23_prior}, but for two different possible {\it{true}} 
 choices of $\tc$. 
 We see that the contours remain qualitatively similar when different {\it{true}} octants of $\tc$ are considered. 
 For the case of maximal mixing, a slight elongation of the contours are observed, especially along the $\db$ and $\dc$ directions. 
 We have also checked that for inverted hierarchy (IH) of neutrino masses our results 
 remain similar. This is due to the fact that even in presence of a sterile neutrino, the mass hierarchy sensitivity of DUNE, though deteriorates, remains high enough ($\gtrsim 5\sigma$)~\cite{Dutta:2016glq}. 
\section*{Appendix C: Understanding the role of channels to probe $\db$ and $\dc$}
\label{apndx_c} 
\renewcommand{\theequation}{C\arabic{equation}}
\setcounter{equation}{0} 
 \begin{figure}[htb]
 \centering
 \includegraphics[scale=0.47]{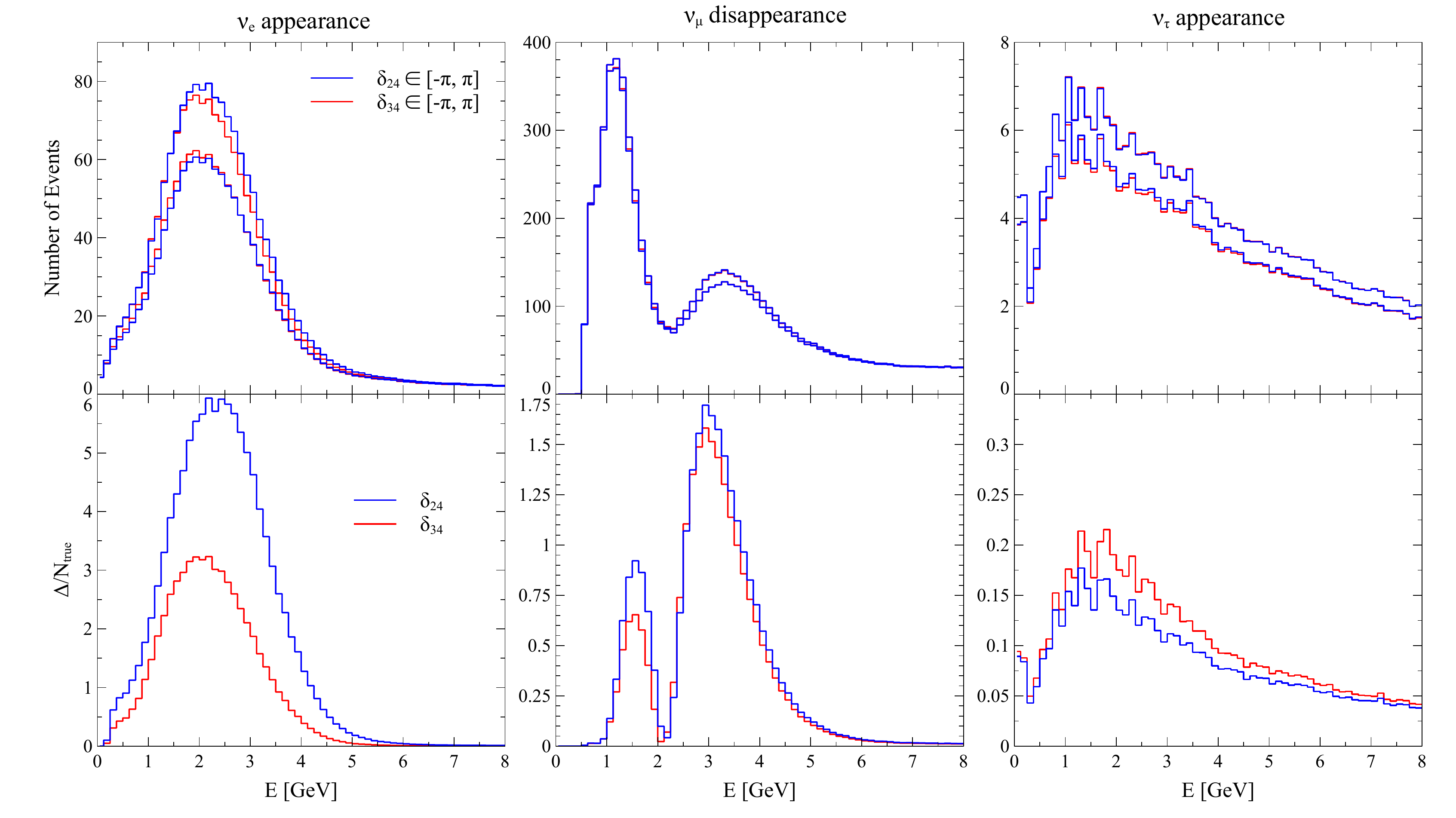}
 \caption{\footnotesize{Top row shows the bands for event spectra at DUNE due to variations of 
 3+1 CP phases for $\nue$ , $\numu$ and $\nutau$  channels respectively. 
 The band within blue (red) lines corresponds to the variation of $\db$ ($\dc$) in the range $[-\pi, \pi]$. 
 The bottom row shows the corresponding spread $\Delta$ (square of difference between the two same coloured curves in the top row) over $N_{\text{true}}$ (number of events when $\db, \dc$ are fixed at zero). See text for details. 
 }}
  \label{fig:event}
 \end{figure}
Here we make an attempt to understand the relative role of the three oscillation channels to probe the sterile CP 
phases $\db$ and $\dc$ as observed in Fig.\ \ref{fig:d_d_channel}. 
For a simpler and more intuitive explanation we use the following Gaussian $\chisq$ (the Poissonian definition in Eq.\ \ref{eq:chisq} reduces to the Gaussian version for sufficiently large number of events~\cite{Zyla:2020zbs}.):
\begin{equation}
\label{eq:chisq_gauss}
\Delta \chi^{2} \sim 
\underset{\text{test}}{\text{Min}} \sum_{\text{bin}}\Bigg[
\frac{\Big(
N_{\mu e}^{\text{true}} - N_{\mu e}^{\text{test}}
\Big)^{2}}{N_{\mu e}^{\text{true}}}
+ \frac{\Big(
N_{\mu \mu}^{\text{true}} - N_{\mu \mu}^{\text{test}}
\Big)^{2}}{N_{\mu \mu}^{\text{true}}}
+ \frac{\Big(
N_{\mu \tau}^{\text{true}} - N_{\mu \tau}^{\text{test}}
\Big)^{2}}{N_{\mu \tau}^{\text{true}}}
 \Bigg] 
 + \text{(prior \& systematics)}.
\end{equation}
Here $N_{\alpha\beta}^{\text{true}}$ ($N_{\alpha\beta}^{\text{test}}$) is the {\it{true}} ({\it{test}}) set of events coming from the $\nu_{\alpha} \to \nu_{\beta}$ oscillation channel. 
Note that, we have sketched here only the relevant statistical part of the $\chisq$ and ignored the 
prior and systematics.  
In order to understand Fig.\ \ref{fig:d_d_channel}, we first estimate the band of $N_{\alpha\beta}^{\text{test}}$ corresponding to individual variations of $\db \in [-\pi, \pi]$ and $\dc \in [-\pi, \pi]$. 
 These spectra are illustrated for the three channels in Fig.\ \ref{fig:event} (top row).  
The $\chisq$ for reconstructing $\db$ and $\dc$ is governed by the spread/width of such event 
bands and we calculate the square of the vertical width at each energy bin and refer it by $\Delta$.
In accordance with Fig.\ \ref{fig:d_d_channel} we then generate $N_{\alpha\beta}^{\text{true}}$ corresponding to $\db = 0$ and $\dc = 0$. 
Now, following Eq.\ \ref{eq:chisq_gauss}, we plot $\Delta/N_{\alpha\beta}^{\text{true}}$ for the 
three channels in the bottom row of Fig.\ \ref{fig:event}, both for $\db$ (blue) and $\dc$ (red). 
This approximately gives us an idea of the relative contribution of the channels in the reconstruction of 
the sterile CP phases.  
As expected, the $\nue$ channel gives the dominant contribution and $\db$-reconstruction is 
expected to be slightly better than $\dc$. 
 $\numu$ channel also plays a role and this mainly comes due to the large number of $\nu_{\mu}$ events. 
 Finally $\nutau$ channel has a small but non-negligible contribution. 
 Both $\numu$ and $\nutau$ channels provide similar capability of reconstructing $\db$ and $\dc$. 
 
  \begin{figure}[htb]
 \centering
 \includegraphics[scale=0.47]{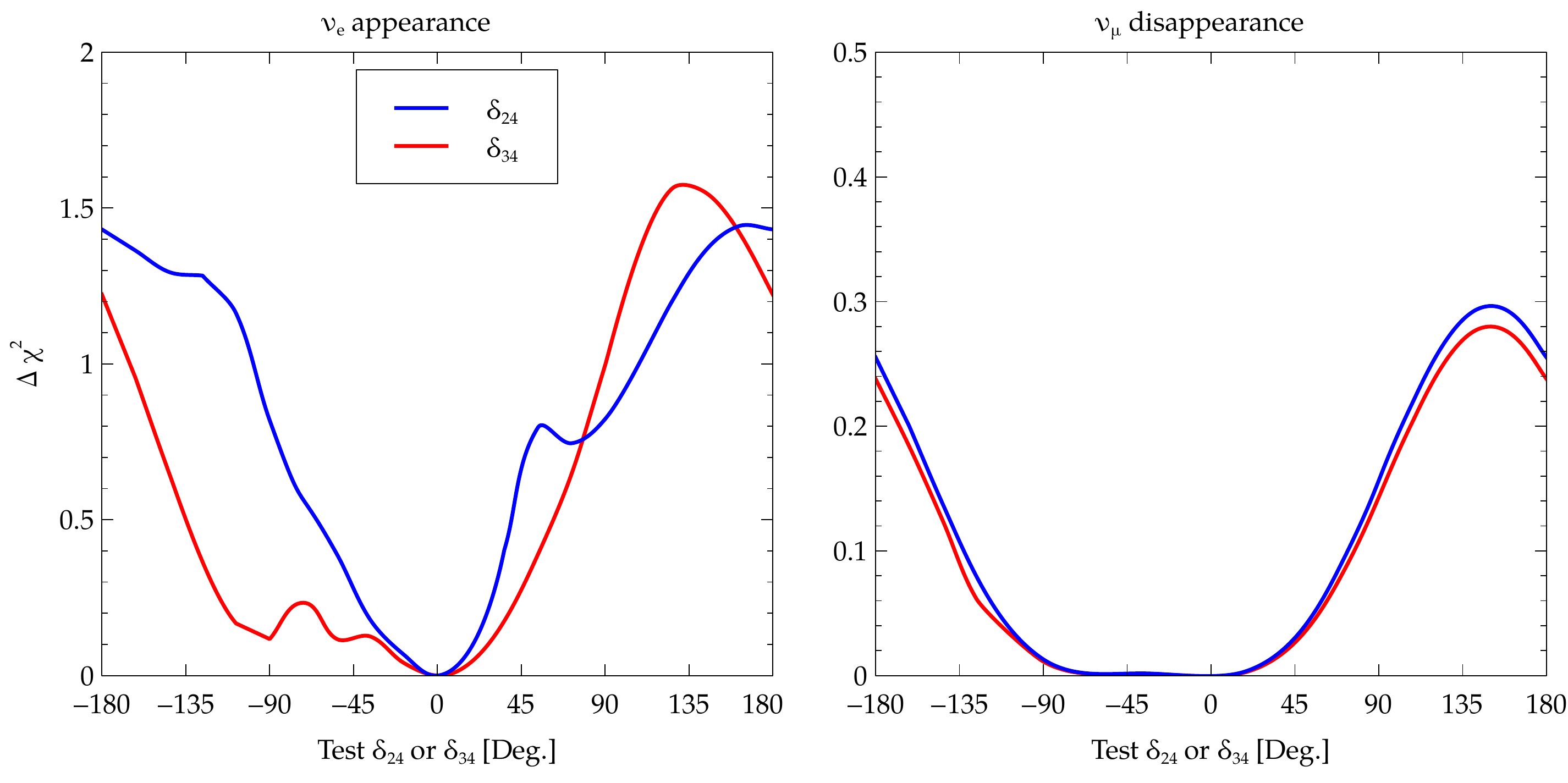}
 \caption{\footnotesize{$\chisq$ in probing $\db$ (blue) or $\dc$ (red) for $\nue$ (left) and $\numu$ (right) 
 channels. The true values of $\db$ or $\dc$ are considered to be zero while the test values  are shown along the horizontal axes.  
 }}
  \label{fig:chisq_syst}
 \end{figure} 
However, since the $\numu$ channel events are also limited by systematic uncertainties, some remarks regarding the effect of systematics in Eq.\ \ref{eq:chisq_gauss} are in order. 
 Extending our analysis in this appendix from the level of events to that of $\chisq$, we show in 
 Fig.\ \ref{fig:chisq_syst}, a comparison between the $\nue$ and $\numu$ channels in probing $\db$ 
 or $\dc$. 
As in Fig.\ \ref{fig:event}, here we also consider the true values of $\db$ or $\dc$ to be 0 and 
then plot the $\chisq$ as a function of the test values of $\db$ (blue) or $\dc$ (red), after marginalising 
over all other relevant test parameters. 
The left (right) panel shows the contribution of the $\nue$ ($\numu$) channel alone. 
 As mentioned in Sec.\ \ref{sec:analysis}, $2\%$ ($5\%$) systematics has been considered in the $\nue$ ($\numu$) signal, following \cite{Alion:2016uaj} (along with various background systematics as well). 
 Due to this larger systematic uncertainty, we note from Fig.\ \ref{fig:chisq_syst}, that the relative role of $\numu$ channel in probing the sterile phases with respect to $\nue$ channel further diminishes compared to 
 that of Fig.\ \ref{fig:event}. 
 Nevertheless, even with such systematics the contribution of the $\numu$ channel can still be 
 roughly $\sim 15\%-20\%$ of that of $\nue$ in the favourable region of the parameter space. 
 Thus when all the oscillation channels are included to do a combined $\chisq$ analysis, the interplay and 
 complementarity among the channels improves the sensitivity to probe the CP phases. 
 These findings are consistent with the full statistical analysis as discussed in Sec.\ \ref{sec:channel}.

\bibliographystyle{apsrev}
\bibliography{reference}
\end{document}